\documentclass[iop,apj]{emulateapj}
\usepackage{amsmath,amssymb,amstext}
\usepackage{float}

\usepackage[breaklinks,colorlinks,citecolor=blue,linkcolor=blue]{hyperref}
\usepackage[all]{hypcap} 
\usepackage{graphicx}
\usepackage{multirow}

\begin{document}

\title{Is Flat-fielding safe for precision CCD astronomy?}

\author{Michael Baumer}
\author{Christopher P. Davis}
\author{Aaron Roodman}

\affil{Kavli Institute for Particle Astrophysics and Cosmology} 
\affil{SLAC National Accelerator Laboratory, Menlo Park, CA 94025, USA}
\affil{Physics Department, Stanford University, Stanford, CA 94305, USA}


\begin{abstract}

The ambitious goals of precision cosmology with wide-field optical surveys such as the Dark Energy Survey (DES) and the Large Synoptic Survey Telescope (LSST) demand, as their foundation, precision CCD astronomy. This in turn requires an understanding of previously uncharacterized sources of systematic error in CCD sensors, many of which manifest themselves as static effective variations in pixel area. Such variation renders a critical assumption behind the traditional procedure of flat fielding---that a sensor's pixels comprise a uniform grid---invalid. In this work, we present a method to infer a curl-free model of a sensor's underlying pixel grid from flat field images, incorporating the superposition of {\it all} electrostatic sensor effects---both known and unknown---present in flat field data. We use these pixel grid models to estimate the overall impact of sensor systematics on photometry, astrometry, and PSF shape measurements in a representative sensor from the Dark Energy Camera (DECam) and a prototype LSST sensor. Applying the method to DECam data recovers known significant sensor effects for which corrections are currently being developed within DES. For an LSST prototype CCD with pixel-response non-uniformity (PRNU) of $0.4\%$, we find the impact of ``improper" flat-fielding on these observables is negligible in nominal $.7''$ seeing conditions. These errors scale linearly with the PRNU, so for future LSST production sensors, which may have larger PRNU, our method provides a way to assess whether pixel-level calibration beyond flat fielding will be required.

\end{abstract}


\section{Introduction}
\label{sec:intro}

Current and next-generation wide-field optical surveys, including the ongoing Dark Energy Survey (DES) \citep{des,des2} and the upcoming Large Synoptic Survey Telescope (LSST) \citep{scibook} will use a variety of cosmological probes, including gravitational lensing, galaxy clusters, and supernovae, to constrain the nature of dark matter and dark energy with unprecedented precision. Though each cosmological probe differs in its priorities, all of them demand, at a fundamental level, ``precision CCD astronomy," placing stringent requirements on instrumental systematics affecting photometry, astrometry, and object shape measurement. These requirements are driving a search for previously unknown sources of systematic error within the CCD sensors used in both the Dark Energy Camera (DECam) \citep{decam} and LSST.

Several such previously uncharacterized sensor systematics have recently been described in the literature. \cite{plazas} analyzed `tree rings'---impurity gradients seen as circularly symmetric flux variations in flat field images---in DECam and found a significant impact on astrometry. \cite{okuraTR} analyzed tree rings in prototype LSST sensors, where due to improved environmental stability during the growth of the parent silicon boules, the tree rings are highly suppressed, and found no significant impact on the two-point shear correlation function. Other effects that have been analyzed in isolation include distortions near CCD edges \citep{bradshaw}, periodic row/column size variations \citep{newyuki}, and stochastic pixel-to-pixel size variation \citep{me!}. 

This previous work has taken a ``bottom-up" approach to assessing the science impact of sensor systematics---typically using pre-processing to highlight a particular effect within the (evolving) taxonomy of uncharacterized errors and assessing the impact of that individual effect on particular science observables. Such analyses have certainly improved our understanding of the ways in which CCDs can lead to systematic errors, but further insight can be gained by considering each of these systematics to be a different characteristic variation of effective pixel areas. With this perspective, the astronomer's primary concern becomes the impact of the superposition of all these sensor effects (along with effects the remain individually undescribed) on astronomical observables. This emphasis on superposition is further motivated by the interpretation of all sensor effects as being caused by spurious lateral electric fields produced by impurity gradients within the silicon bulk of CCD sensors, as suggested in \cite{smith}, and described further in \cite{holland}, \cite{lupton}, and \cite{stubbs}. 

When pixel areas vary across a CCD sensor, a fundamental assumption underlying the near-ubiquitous practice of flat-fielding astronomical images is violated (to some degree). Flat fielding assumes that a sensor's pixel grid is perfectly uniform, and that the entirety of its pixel response non-uniformity (PRNU), measured as the standard deviation of a luminosity-corrected coadded flat field, is caused solely by local variations in the quantum efficiency (QE) of the sensor. In this case, dividing a raw image by a flat field image yields a calibrated flux in each pixel, as is desired. However, if sensor effects cause the pixel grid to deviate from perfect rectilinearity, dividing raw science image pixel values by flat field pixel values yields a calibrated flux per unit (unknown) area---a ``pixel surface brightness" rather than a flux---for each pixel in the image. Since astronomical image analyses typically assume that each pixel value represents a calibrated flux, it is critical to understand the extent to which deviations from a rectilinear pixel grid break this fundamental assumption and potentially render flat-fielding inappropriate. 

In this work, we present a novel method for inferring a model of a sensor's underlying pixel grid from coadded flat-field images, with results from applying the method to representative DECam and LSST prototype sensors. In contrast to previous work in this area, we take a ``top-down" approach by considering the superposition of all static sensor effects simultaneously by inferring a pixel grid model directly from minimally-processed flat-field data. There are several advantages to this method. The first is that rather than requiring a complex sequence of calibration exposures, the only data required are several hundred flat-field exposures, likely already on hand at most observatories and sensor laboratories. Secondly, by analyzing flat-field data with minimal pre-processing, we remain sensitive to potential new, as yet uncharacterized, sensor effects and unbiased by the pre-processing that is typically required to highlight a particular effect of interest. Finally, with our top-down approach, we can test the science impact of all effects at once, allowing us to account for possible constructive and/or destructive interference between the different sensor effects characterized in previous work.

Since our goal is to place an upper bound on the impact of the ``improper" flat-fielding described above, we will assume in this analysis that the assumptions of flat-fielding are maximally wrong---i.e. that {\it all} of a sensor's PRNU is due to the distortion of the underlying pixel grid rather than variations in sensor QE. For a given sensor, if this upper bound is still within the science requirements, then flat-fielding is an acceptable method of calibrating pixel fluxes. \cite{plazas} have already demonstrated the need for pixel-level corrections of DECam astrometry to account for tree rings; a central goal of this paper will be to determine if similar pixel-level corrections beyond flat-fielding will be necessary for precision astronomy, and therefore cosmology, with LSST. 

Having motivated this work and described its fundamental assumptions, we describe our method for fitting a model of an underlying pixel grid to flat field data and assessing the impact of deviations from grid rectilinearity on measurements of simulated PSFs in section \ref{sec:method}. In section \ref{sec:results} we validate the results of our grid model fits on both simulations and data, and present the results of our science impact analysis for a representative DECam and low-PRNU LSST prototype sensor. Section \ref{sec:discussion} discusses the implications of these results, including their generalization to future production-quality LSST sensors, and section \ref{sec:conclusions} outlines potential future work informed by these results.

\section{Methods}
\label{sec:method}

Before fitting an underlying grid model to flat-field data, a reasonable parametrization must be chosen for the underlying pixel grid distortions. Full electrostatic simulations such as those of \cite{andy} can compute pixel boundary shapes with arbitrary precision, but require representing the CCD sensor, including defects in the silicon bulk, as an underlying charge distribution that cannot be precisely known for each individual device. 

Though electrostatic simulations have shown that pixel boundaries can have complex geometries \citep{andy}, they remain approximately linear in the limit of weak pixel grid distortions. This constraint substantially reduces the number of parameters required to describe the pixel grid, relative to a full electrostatic simulation. We therefore parametrize pixel grid models by vertex positions, demanding linear pixel boundaries between displaced vertices.

\begin{figure}[t]
    \centering
    \includegraphics[width=\columnwidth]{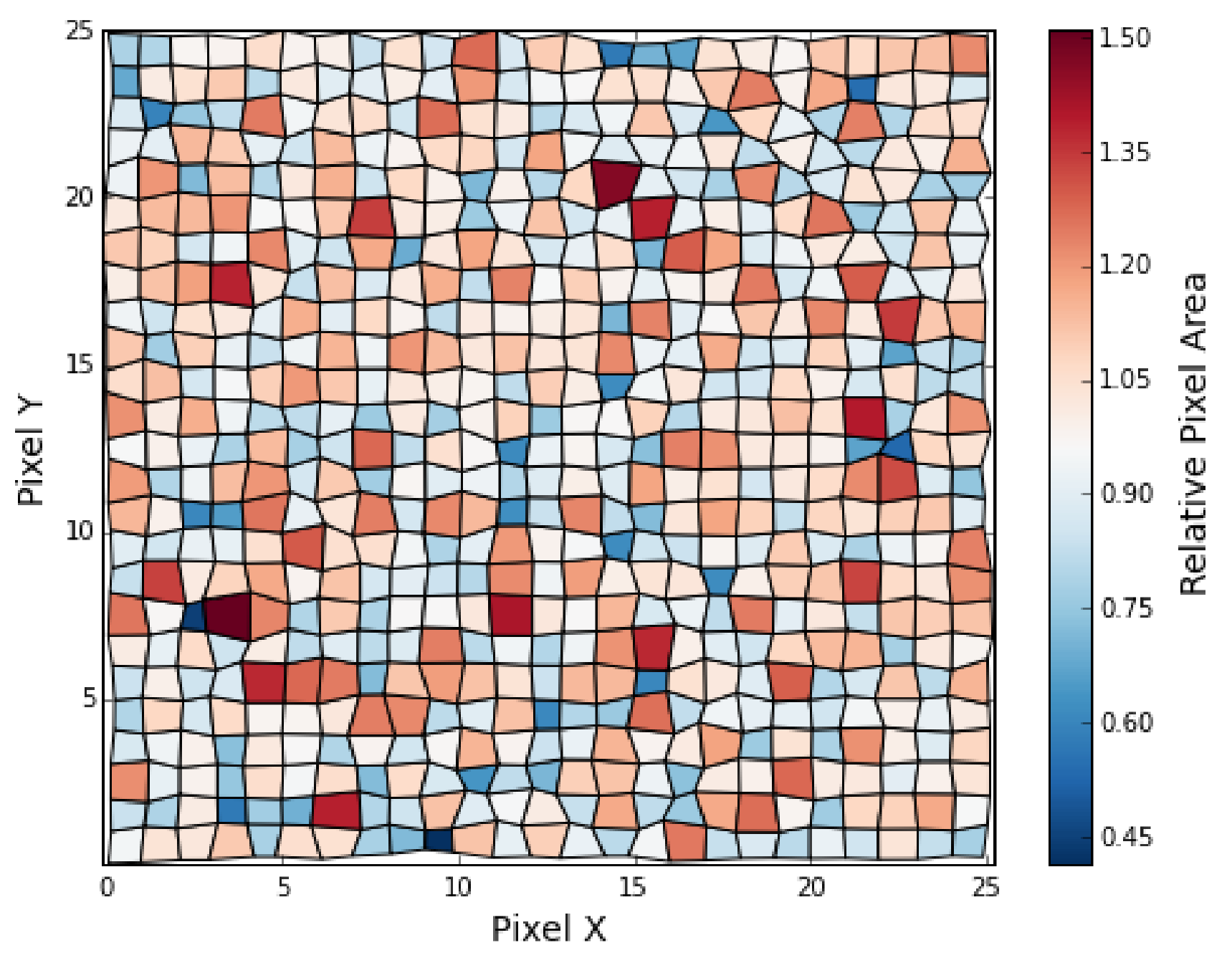}
    \caption{\normalsize A $25 \times 25$ cutout of a fitted pixel grid model of an LSST sensor, with vertex perturbations magnified by a factor of 40 to make grid distortions visible by eye. Pixel color indicates pixel area relative to a unit rectilinear grid.}
    \label{fig:cartoon}
\end{figure}

A $25 \times 25$ cutout from a fitted pixel grid model of an LSST sensor is shown in Figure \ref{fig:cartoon}, with vertex displacements scaled up by a factor of 40 to make the grid distortions (with their characteristic pixel-neighbor anti-correlations as observed in \cite{me!}) readily visible. 

\hspace{1.25in}

With this model of pixel area distortions, we analytically compute the area of a distorted pixel using a standard cross-product formula for the area of a convex quadrilateral: 

\begin{align}
    \hat{A}_{ij} = \frac{1}{2} \big| &(x_{i+1,j+1} - x_{i,j}) (y_{i,j+1} - y_{i+1,j}) \notag\\ - &(x_{i,j+1} - x_{i+1,j}) (y_{i+1,j+1} - y_{i,j}) \big|
    \label{eqn:area}
\end{align}

When optimizing a pixel grid model, we desire these computed model areas to match the observed areas (from the coadded flat fields) as closely as possible. Since equation \ref{eqn:area} gives the area of any vertex configuration, this allows us to define a goodness-of-fit statistic for a single pixel as the square of the difference between the computed model area and area in the data. By summing over all pixels, we can construct a cost function of the form:
\begin{align}
    \mathcal{C} = \sum \limits_{\text{pixels } i,j} (\hat{A}_{ij}(\text{vertices}) - A_{ij})^2
    \label{eqn:cost}
\end{align}
where $\hat{A}_{ij}$ is the computed area of the inferred vertex arrangement at pixel (i,j), and $A_{ij}$ is the true area of the pixel, as measured by the flux in a coadded flat field. 

To minimize this cost function via gradient descent, we need to calculate the derivative of $A_{ij}$ with respect to the positions of its four neighboring vertices. These derivatives can be calculated analytically for each vertex using equation \ref{eqn:area}. This gives an overall gradient of $\mathcal{C}$ with respect to the positions of each vertex. The gradient descent update rule for each vertex position $\vec{x}_{ij}$ is then given by

\begin{align}
\vec{x}_{ij} \rightarrow \vec{x}_{ij} - \lambda \frac{d\mathcal{C}}{d{\vec{x}_{ij} }}
\end{align}
where $\lambda$ is a tuning parameter that sets the step size of the parameter update along the gradient direction, and the negative sign ensures the update is opposite the gradient direction, leading to a decrease in the cost function $\mathcal{C}$. The value of $\lambda$ must be tuned empirically via a grid search to optimize model convergence. Code implementing this fitting procedure, built in Python around the \verb+numpy+/\verb+scipy+ \citep{numpy}, \verb+matplotlib+ \citep{matplotlib}, and \verb+pandas+ \citep{pandas} libraries, is made available at \url{https://github.com/cpadavis/weak_sauce}. Demonstrations of the code are also provided as IPython notebooks \citep{ipython}. 

We use the following technique to assess the impact of the pixel grid distortions inferred by these fits on critical astronomical observables. We use GalSim \citep{galsim} to construct flux profiles for Moffat PSFs with a nominal seeing value ($0.9"$ for DECam and $0.7"$ for LSST). Surface integration of these flux profiles over each pixel of the distorted grid is performed using a two-dimensional adaptation of a code designed for exact voxelization of phase space in N-body simulations \citep{devon}. This results in a rendering of the true flux profile as `observed' by the distorted pixel grid of the sensor. This rendered image is then divided by the coadded flat field, mimicking a standard data reduction pipeline. In this case, however, because the PRNU has been modeled as pixel grid distortions rather than QE variation, this flat fielding does not properly correct the raw image.

Finally, we use an implementation of the adaptive image moments algorithm described in \cite{hsm} to compare observed parameters of sources rendered onto both our fitted pixel grid and an ideal rectilinear grid. This algorithm computes image moments using an adaptive elliptical weight matrix. This allows the impact of static pixel grid distortions on photometry (zeroth moment), astrometry (first moment), and shape measurements (second moments) to be evaluated. We use the following (unnormalized) convention for ellipticity to allow use of $e_0$ as a proxy for PSF size:

\begin{align}
e_0 = I_{xx} + I_{yy} \;\;\;\;\;\;\;
e_1 = I_{xx} - I_{yy}\;\;\;\;\;\;\;
e_2 = 2I_{xy}
\end{align}
where $I_{ij}$ are weighted second moments of the rendered source image.

\subsection{Data}
\label{sec:data}

A major advantage of this method of assessing sensor systematics is the simplicity of obtaining the required data. While star flats, for example, are a common way of assessing the impact of sensor imperfections on astrometry, they require the sensor to be mounted on a telescope where time is available for a large number of dithered exposures. In the lab, dithered arrays of pinholes can be imaged to allow for similar measurements, but require a complex laboratory setup. 

In contrast, the method presented in this paper requires only flat field images, which are routinely acquired in large numbers in both sensor laboratory and observatory settings. To suppress the shot noise that dominates the variance of individual exposures, we coadd $\sim 500$ flat field images taken with an LSST prototype CCD at a wavelength of $625 \text{nm}$ and a light level of 75,000e$^-$. Each frame is bias- and overscan-corrected and rescaled to the same average flux level before co-adding to account for shutter timing variability. For DECam, a similar number of r-band dome flats are coadded, with identical pre-processing. In both cases, we consider a single CCD amplifier block (1/16 of a chip for LSST; 1/2 chip for DECam) for simplicity, cropping chip edges to focus on effects in the central regions which provide the bulk of quality wide-field survey data. Bad pixels were masked both in the cost function optimization (to avoid warping their neighboring pixels) and the systematics analysis.

We subtract a seventh order polynomial fit from the coadded flat fields to remove variation in the illumination pattern. This step has the additional benefit of removing large-scale flux variations that are predominantly due to QE variations. A primary determinant of the QE of a pixel is the quality of the anti-reflective (AR) coating on the pixel surface. Fluctuations in AR coating quality across a sensor result in positive pixel-neighbor correlations. As is the case with illumination variation, if a pixel is brighter (dimmer) than nominal due to fluctuations in AR coating quality, the relatively large scale of AR coating variations relative to the pixel scale means that its neighbors will likely also be brighter (dimmer). 

On the other hand, local pixel grid distortions are caused primarily by imperfections in the silicon bulk. These lead to negative pixel-neighbor correlations as pixels gain and lose area from their neighbors (as shown in \cite{me!}), which are not removable with a polynomial illumination correction. The fact that the large-scale variations removed by this illumination correction are caused primarily by QE variations across the sensor only increases the validity of our worst-case assumption that all remaining PRNU is due to distortions of the underlying pixel grid.

In the following section, we validate the methods presented in the section using both simulations and data, and present results obtained from applying the method to both the DECam and LSST data described in this section.

\section{Results}
\label{sec:results}

\subsection{Validation}
\label{sec:validation}

One potential concern in fitting this model to data is that it is apparently underconstrained by a factor of $\sim 2$, with $N \times M$ data points from the coadded flat fields and $2 \times (N+1) \times (M+1)$ parameters from allowing each vertex to shift in two dimensions. Expanding this out, we can interpret the $2NM$ term as constraining the pixels in the bulk of the sensor, the $2N+2M$ terms as constraining the boundary axes, and the trailing 2 degrees of freedom as a global translational shift. 

Consider the grid of vertex displacements as a vector field, which can be represented as the sum of a pure divergence and a pure curl term. In this context, a divergence of pixel vertices (`E-mode') around a given pixel changes that pixel's area, while a curl (`B-mode') does not. Since the optimization of our cost function is an attempt to match the pixel areas given by the model with the pixel areas implied by the flat field, the optimization process will only introduce curl-free (area-changing) displacements into the pixel grid.

This means that it will not induce an (area-preserving) global translational shift of the sensor grid, leaving these 2 parameters fixed. It will also not slide rows and columns of vertices past each other (which would lead to a B-mode shear of entire rows and columns of pixels), which fixes the $2N+2M$ boundary parameters that control row and column offsets. This leaves $2NM$ degrees of freedom, of which the $NM$ E-mode parameters are well-constrained by the $NM$ data points from the flat field, while the $NM$ B-mode parameters are unconstrained. Since the fitting procedure begins from a rectilinear grid, for which the 'B-mode' amplitude is zero, no B-mode vertex displacements will be present in the resulting fit. The model will always return a pixel grid model with pure E-mode distortions, and therefore will not be able to constrain B-mode distortions that may be present in the pixel grid.

Despite its ability to constrain only E-mode vertex distortions this model is still very useful, since as discussed in Section \ref{sec:intro}, we expect most sensor systematics to result from spurious transverse electric fields in the silicon bulk. Since Maxwell's equations ensure that an electrostatic field will be curl-free, all electrostatic sensor systematics (even those yet uncharacterized) will cause only E-mode distortions in the pixel grid. Such distortions are a perfect match for the capability of our model.

\begin{figure}
    \centering
    \includegraphics[width=0.7\columnwidth]{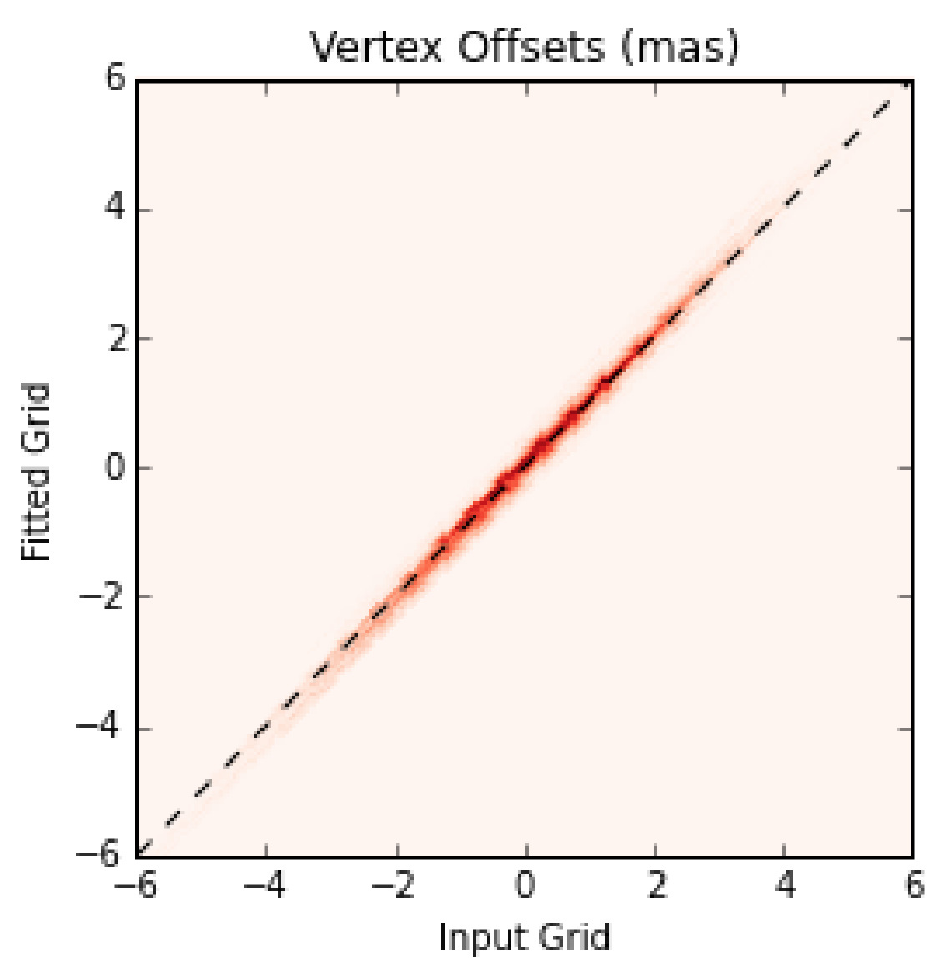}
    \caption{\normalsize When fitting to flat fields from pixel grids with pure E-mode distortions, the model returns the input vertex offsets with high fidelity. This will be the case for all electrostatic (curl-free) sensor systematics.}
    \label{fig:emodes}
\end{figure}

Indeed, when we fit our model to flat fields synthesized from pixel grids with known E-mode distortions (pixel displacements that are the gradient of a scalar field), our fitted model captures the input vertex positions with high fidelity. This is demonstrated in Figure \ref{fig:emodes}, which shows a tight correlation between the known and recovered locations of pixel vertices for sensors with pure E-mode vertex distortions.

\begin{figure}[htbp]
    \centering
    \includegraphics[width=\columnwidth]{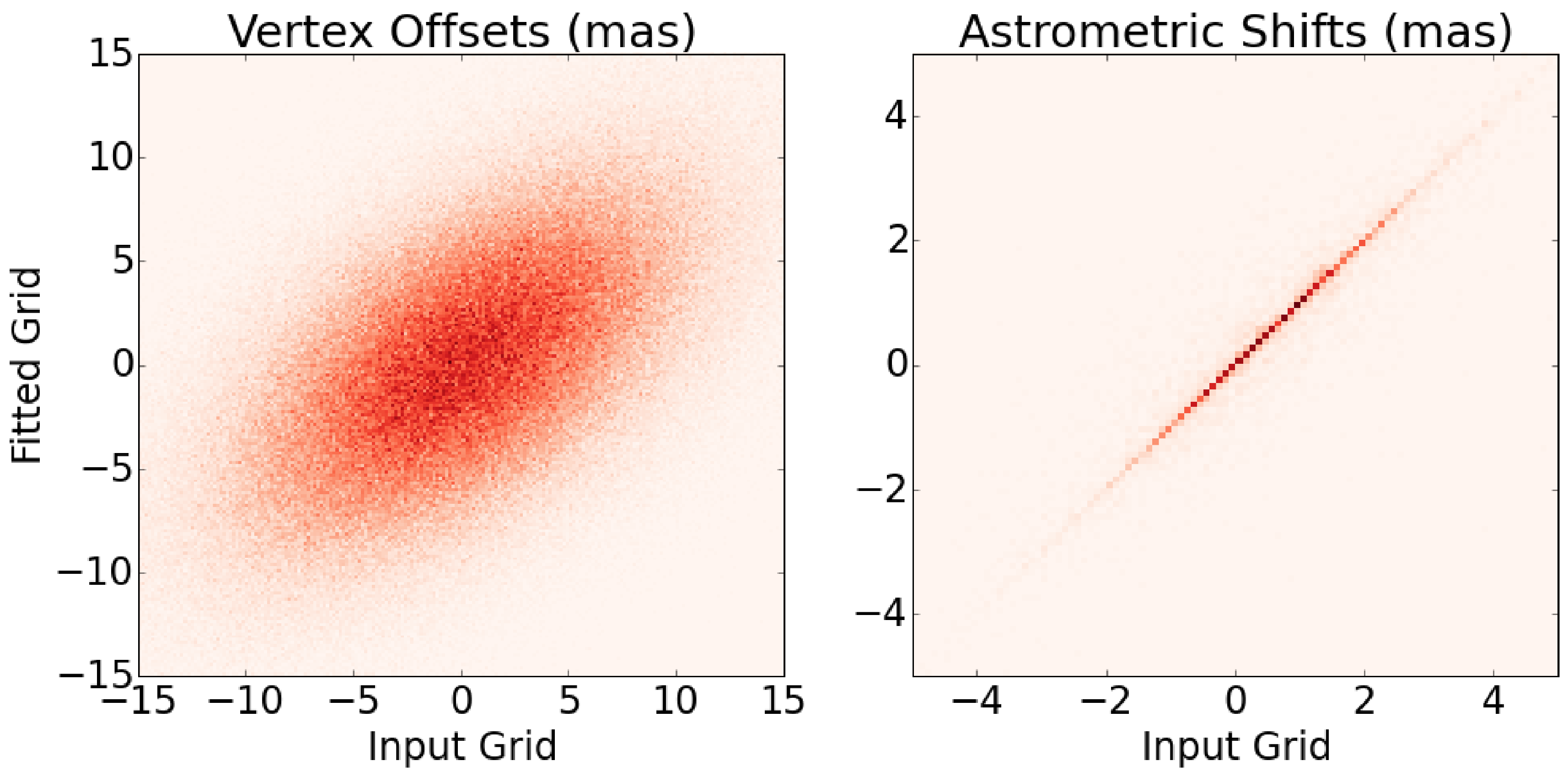}
    \caption{\normalsize For flat fields from pixel grids containing spatially unstructured B-modes, the model's reconstruction of the pixel grid is very noisy. However, since these errors induced by the uncorrelated B-modes average out under PSF profile (FWHM $\sim$ 3.5 pixels), the systematic errors induced by distortions in the input and recovered grids on PSF observables show excellent correlation.}
    \label{fig:sims}
\end{figure}

If B-mode distortions are present in a pixel grid, their impact on the results of this fitting procedure depends on their degree of spatial correlation. For example, when fitting to flat fields from known grids with both E- and B-mode power (constructed by perturbing a rectilinear grid with independently-drawn Gaussian vertex displacements), the recovery of individual vertex positions is degraded (see the left panel of Figure \ref{fig:sims}), since does not capture the B-mode distortions present in the input grid.

However, since the B-modes are spatially uncorrelated in this case, the residuals that result from failing to capture them in the model average out under a PSF profile. The right panel of Figure \ref{fig:sims}, which plots the astrometric shifts measured on the recovered grid against those measured on the input grid, shows that these spatially-uncorrelated uncertainties in individual vertex positions do not affect the analysis of the impact of pixel grid distortions on PSF observables. 

So as long as there are no B-mode distortions that are spatially correlated at large scales, our model will capture the systematic impact of all sensor effects properly. All of the individual effects mentioned in section \ref{sec:intro} (tree rings, edge effects, and periodic step-and-repeat errors from the lithographic mask, and intrinsic pixel-to-pixel size variation) all result in pure E-mode distortions, which means they will all be captured by this fitting method.

One possible source of B-mode distortions is ``lithographic jitter"---pixel-to-pixel imperfections in the lithographic mask. While this effect is not constrained to induce pure E-mode distortions, any B-modes induced by errors in the mask are likely to be spatially uncorrelated, and therefore unlikely to affect PSF observables, as shown in Figure \ref{fig:sims}. We are not aware of any sources of spatially-correlated B-mode distortions of CCD pixel grids. However, should any exist, they would not be constrained by this fitting procedure, and would require separate characterization. In this work, we will proceed under the assumption that all sensor systematics affecting astronomical observables are fundamentally electrostatic (i.e. curl-free), and present further validations of this method on real data.

To demonstrate that our fitted grid is not only a theoretically robust solution, but a physically relevant one, we illustrate the pixel shapes (widths and heights) from a model fit to an LSST sensor with known anisotropy in pixel boundary strengths and DECam sensor with tree rings.

\begin{figure}[!htbp]
    \centering
    \includegraphics[width=\columnwidth]{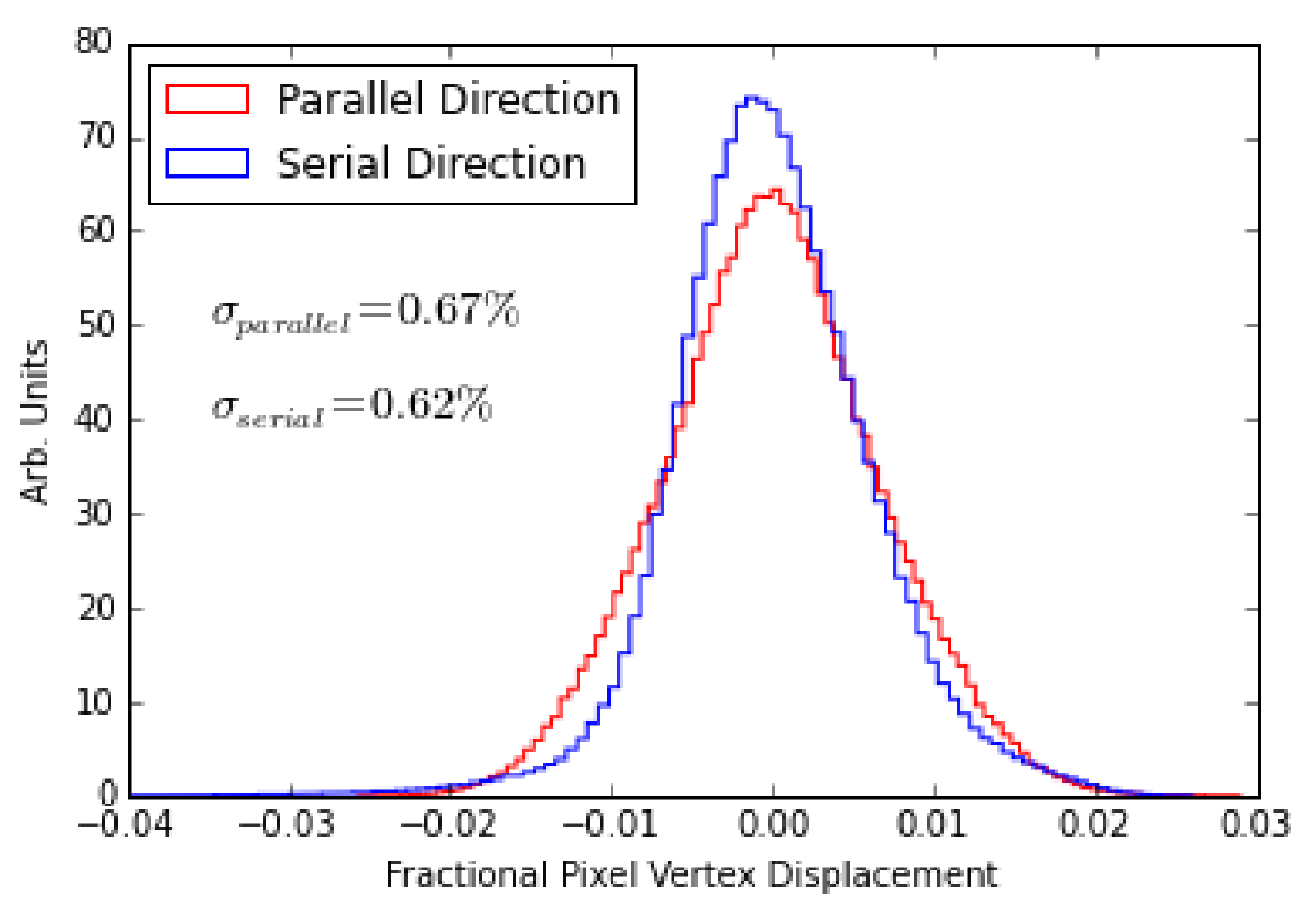}
    \caption{\normalsize A histogram of vertex displacements from the nominal pixel grid, illustrating the relative susceptibility of the weaker clock leaves which serve as the pixel boundaries in the parallel direction to the stronger channel stop implants bounding the serial direction, as predicted in \cite{andy}.}
    \label{fig:vert_hist}
\end{figure}

Figure \ref{fig:vert_hist} illustrates how a pixel grid fit to an LSST coadded flat field is very close to rectilinear---the RMS deviation of pixels from their nominal position is $\sim 1/1000$ of a pixel side length. Given that the vertex displacements remain small in the absence of an explicit regularizing constraint, this figure illustrates directly that explicit regularization is unnecessary. We observe that the scale of the vertex displacements in the parallel direction (vertical) is slightly larger than in the serial direction (horizontal). This is expected behavior from both electrostatic simulations \citep{andy} and data \citep{me!}, where the anisotropy is explained via the relative electrostatic weakness of the pixel clock boundaries (parallel) with respect to the stronger, static channel stop implants (serial).

\begin{figure}[htbp]
    \centering
    \includegraphics[width=.7\columnwidth]{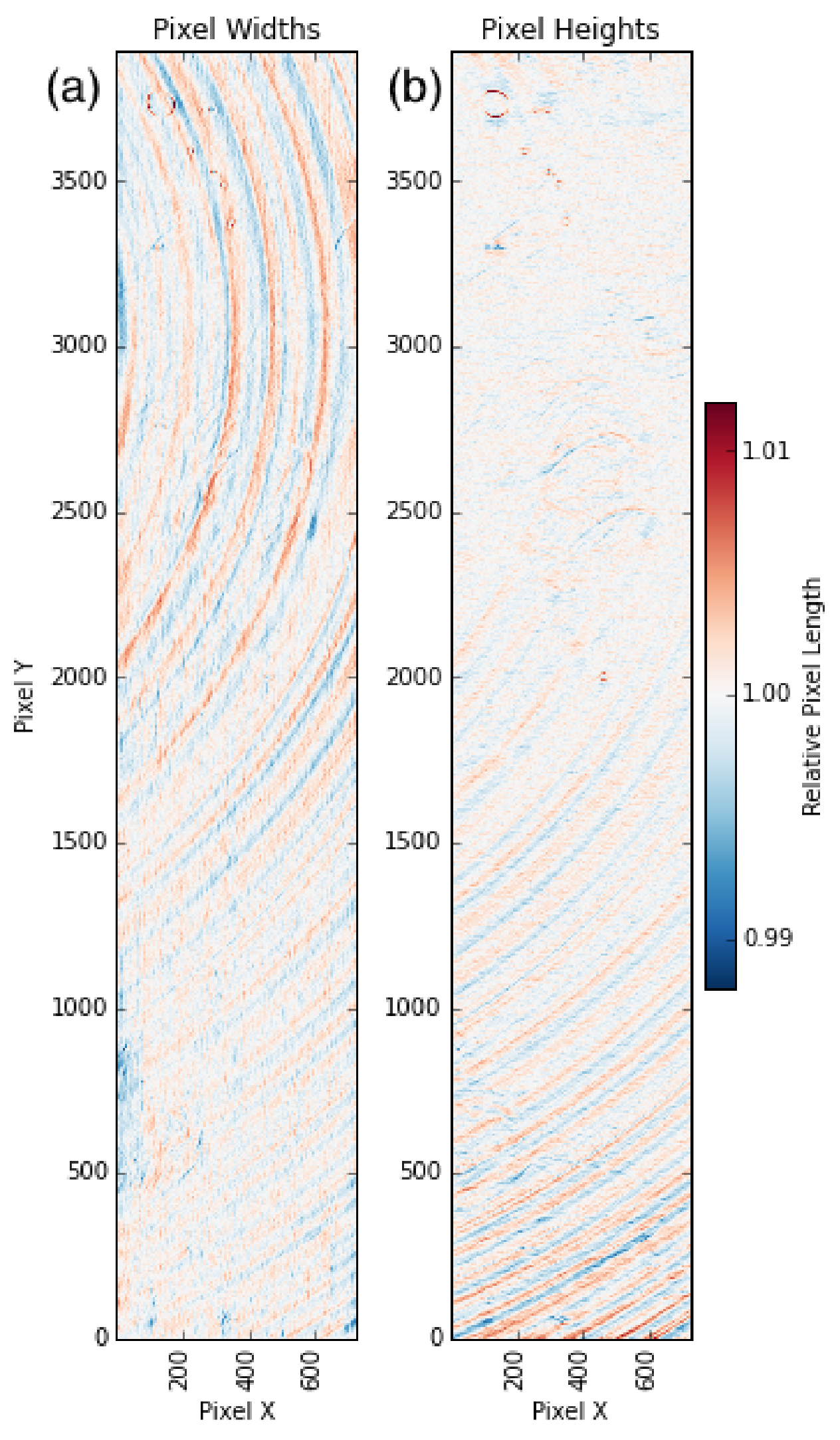}
    \caption{\normalsize Maps of relative pixel widths (a) and heights (b) from a fit to a coadded DECam flat field, showing preferential variation in pixel shape along the radial direction of the tree rings, as expected. This demonstrates that the pixel grid models inferred by our method are physically relevant.}
    \label{fig:pixshapes}
\end{figure}

Figure \ref{fig:pixshapes} illustrates our correct inference of pixel shapes due to a dominant tree-ring distortion in a DECam sensor. Since tree rings are caused by radial transverse electric fields, pixel widths (heights) are expected to be most affected when the tree ring is fluctuating in the horizontal (vertical) direction. This pattern is observed in Figure \ref{fig:pixshapes}, showing that the fitted pixel geometries reproduce the expected pixel shape anisotropy of tree ring features in DECam. This expected behavior is realized in the pixel grid fits despite no explicit model pressure (the model knows only about pixel fluxes, not shapes), demonstrating that this method infers physically reasonable pixel geometries from real data.

\begin{figure}[htbp]
    \centering
    \includegraphics[width=\columnwidth]{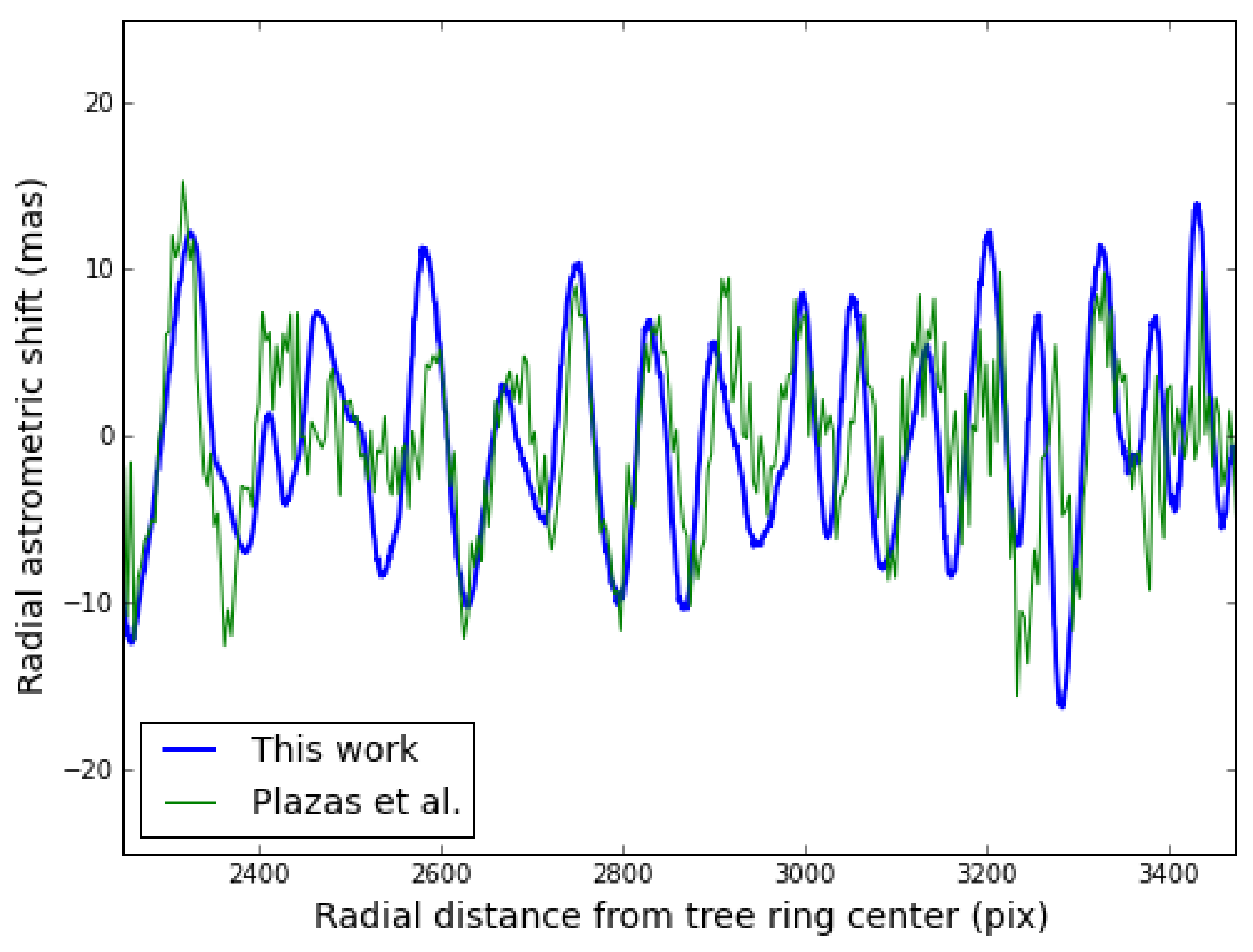}
    \caption{\normalsize Radial profiles of radial astrometric distortion as measured using DECam star flats from \cite{plazas}, compared with the prediction from our pixel grid model, showing good agreement. The radial profile presented in this work is smoother than the one derived from star flats, as our full grid model is derived from coadded flat fields with negligible photometric (shot) noise.}
    \label{fig:andres}
\end{figure}

An even more direct test of this method is illustrated in Figure \ref{fig:andres}, which shows a comparison of the radial profile of radial astrometric distortions as measured using DECam star flats and the prediction from our pixel grid model. The radial profiles of tree rings as measured by our model are well-aligned with those measured by \cite{plazas}. The small differences are likely the results of either noise (since our model, derived from coadded flat fields with negligible photometric noise, can be sampled much more densely than star flats allow on the sky) or tree ring structure being washed out in our model by local QE defects present in DECam sensors (scratches, smudges, etc.) that our method incorrectly models as pixel size variation. This latter concern is not an issue in analyzing the LSST sensor, where the cosmetic quality is extremely high, but should be considered when applying this method to sensors with significant variations in both QE and pixel size.

Having demonstrated the robustness of this method for inferring sensor systematics from flat field data via pixel grid models, we present results from applying our methods to data from a representative DECam CCD and an LSST prototype sensor in the following sections.

\subsection{Results from DECam}
\label{sec:decam}

\begin{figure}[ht]
    \centering
    \includegraphics[width=.7\columnwidth]{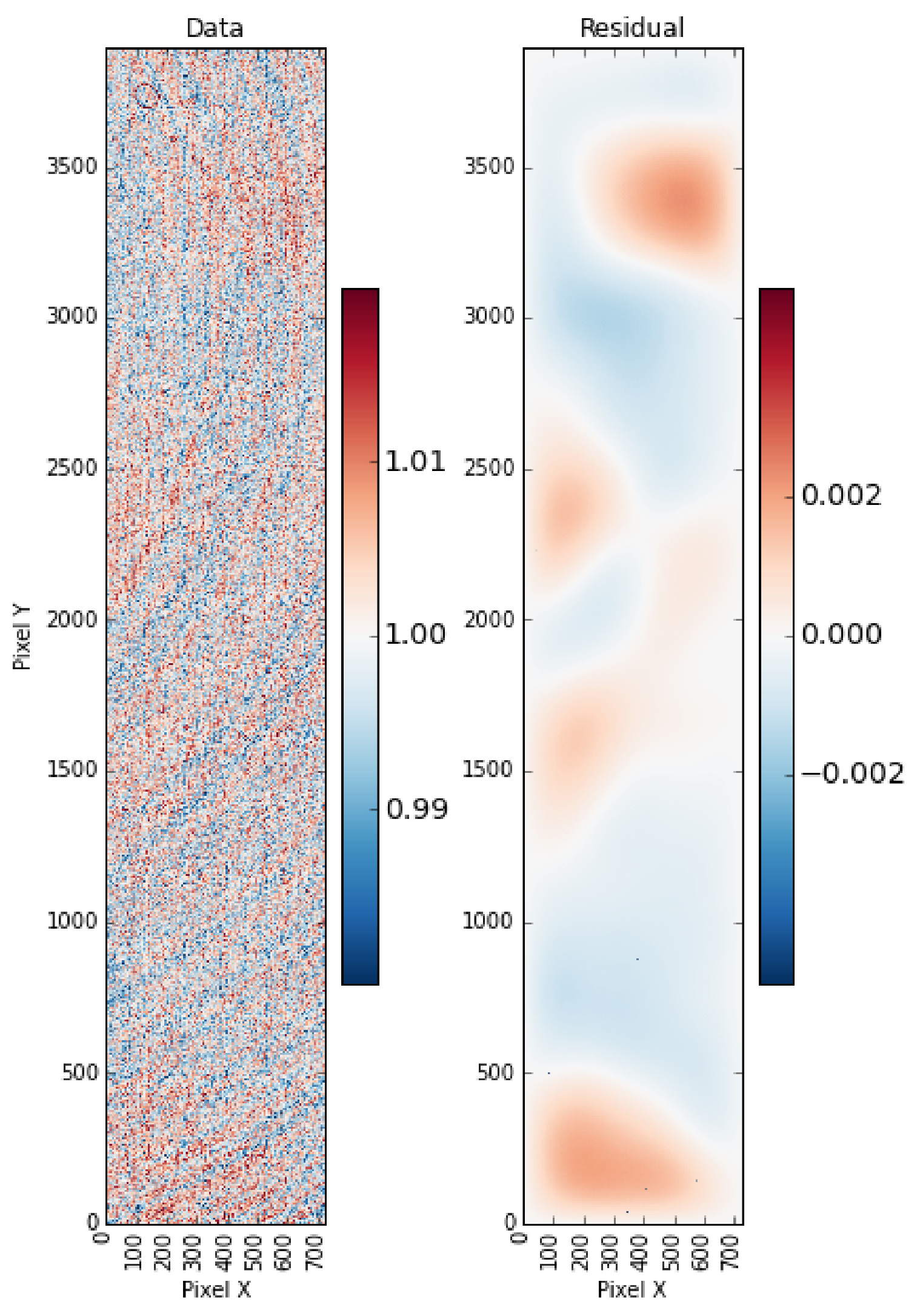}
    \caption{\normalsize At left, a one-amplifier cutout of a coadded DECam flat field, which the pixel grid aims to model during the fit. At right, a map of the flux residuals for the fitted grid model. The lack of similarity between the data and residuals demonstrates that the model is picking up the relevant spatial structure in the flat field, in particular the prominent tree rings.}
    \label{fig:des-resid}
\end{figure}

\begin{figure*}[!htbp]
    \centering
    \includegraphics[height=3in]{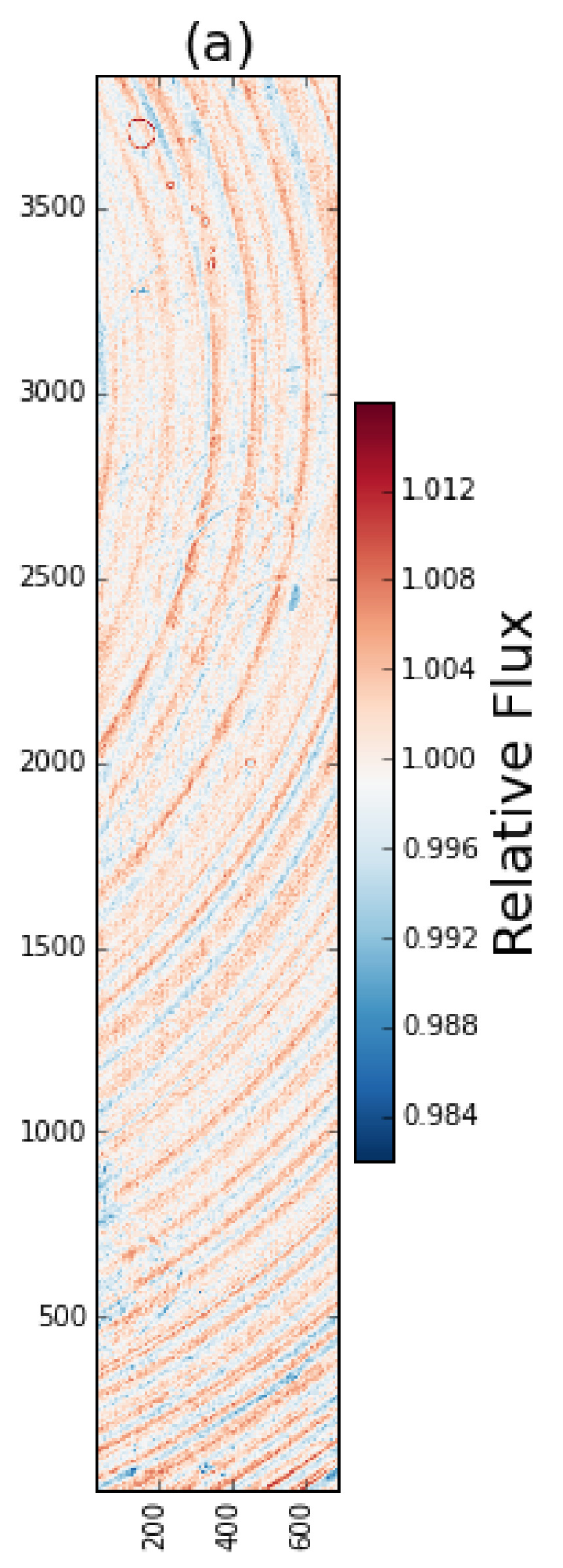}
    \includegraphics[height=3in]{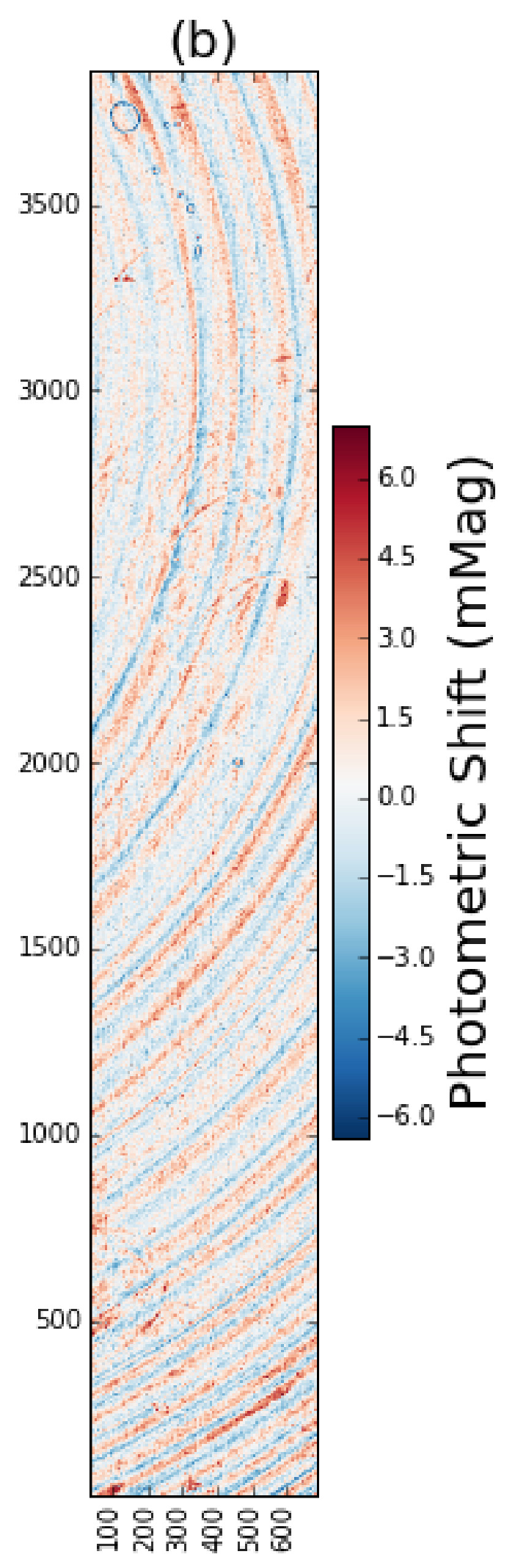}
    \includegraphics[height=3in]{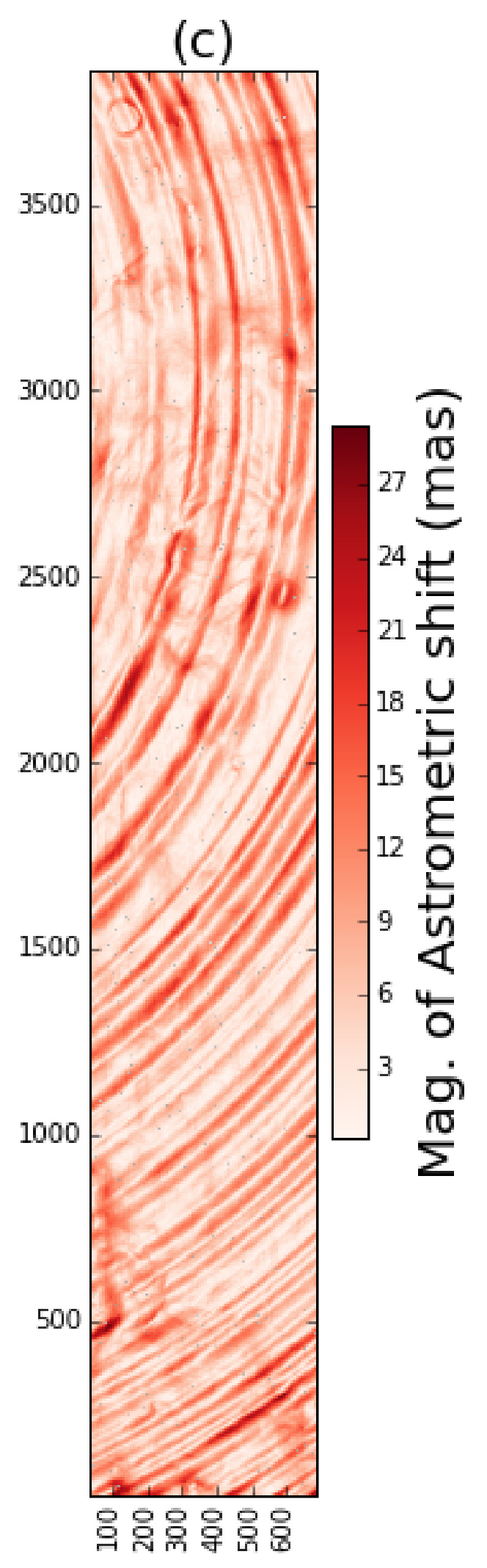}
    \includegraphics[height=3in]{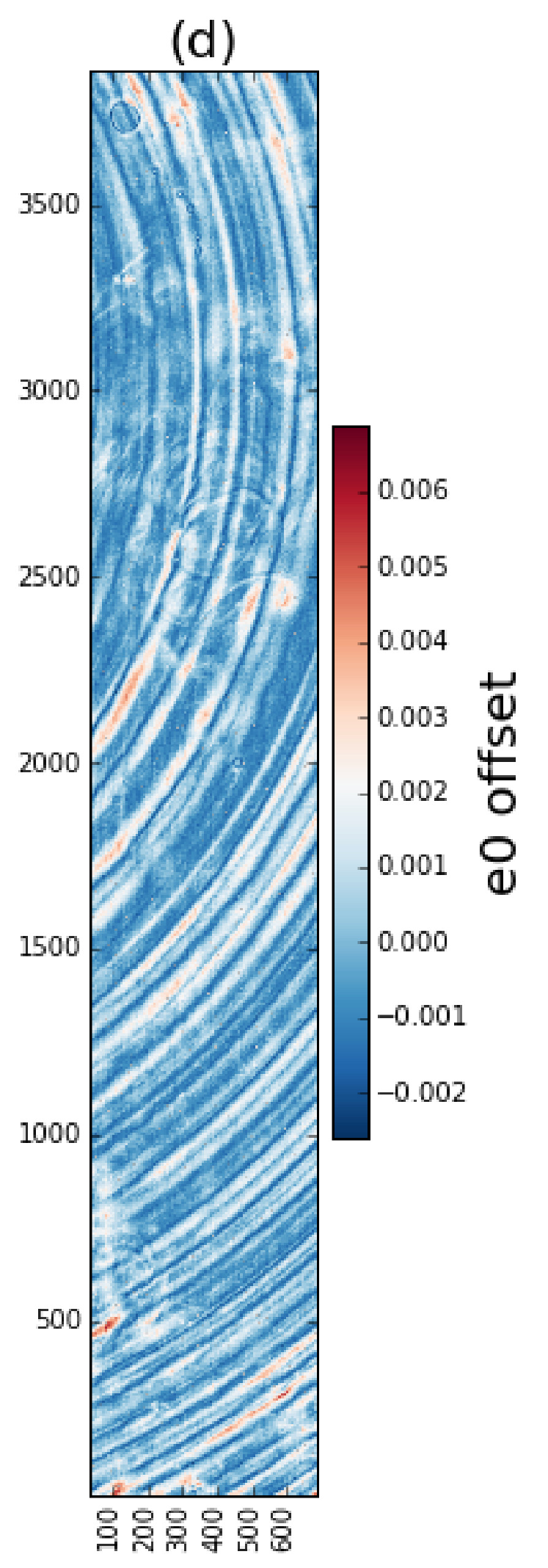}
    \includegraphics[height=3in]{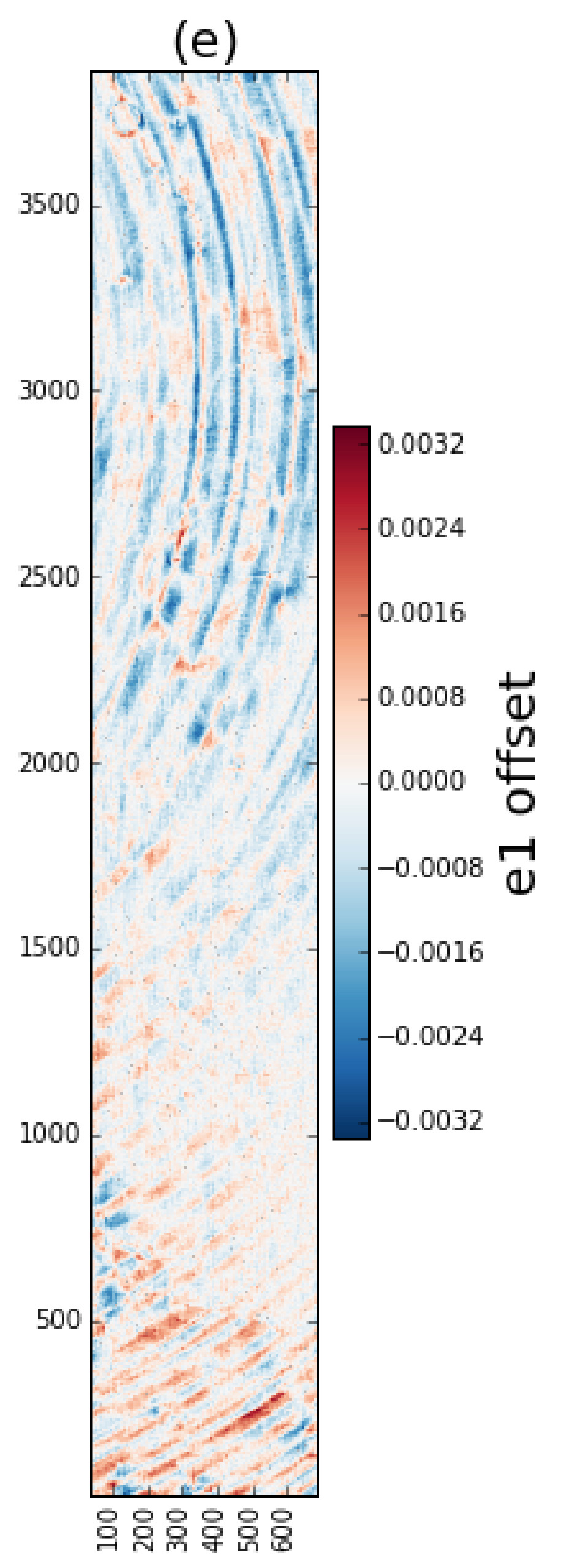}
    \includegraphics[height=3in]{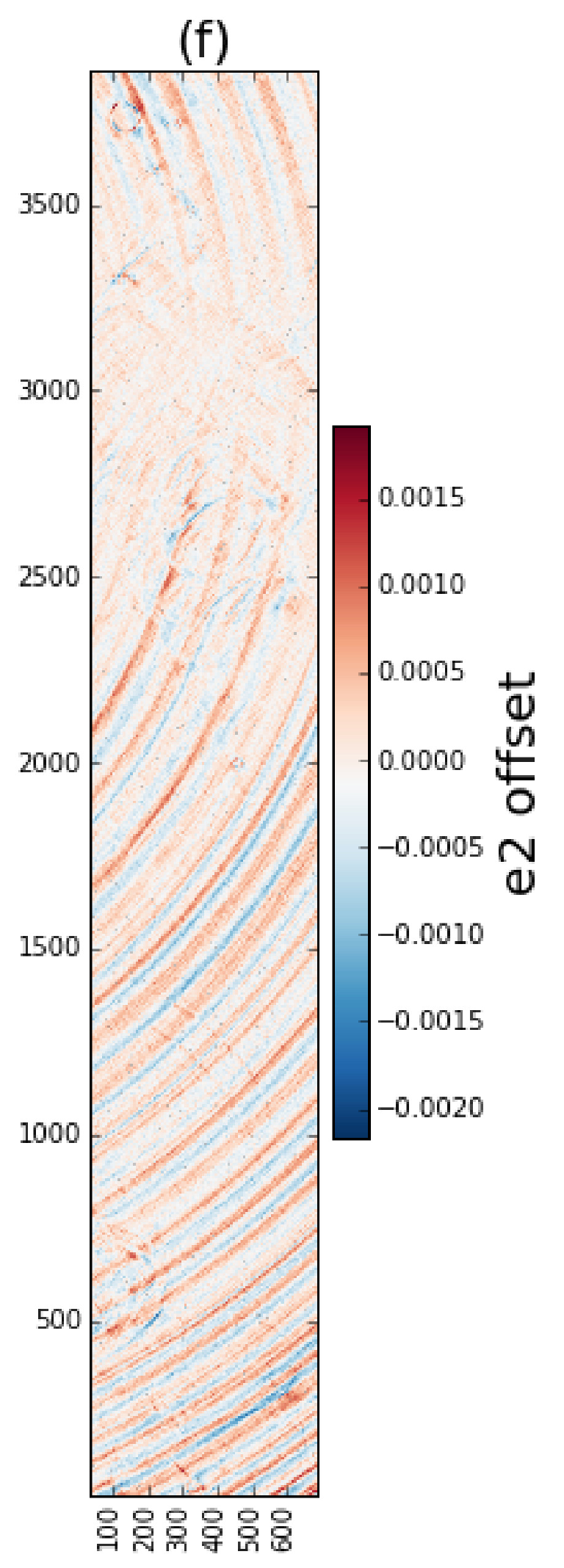}
    \caption{\normalsize A summary of systematic errors induced by sensor effects across a DECam chip, as modeled using our method. A flat field (a) (smoothed to highlight medium-scale structure) for comparison with systematics maps from a representative DECam amplifier, showing significant impact on photometry (b), astrometry (c), PSF size (d), and PSF ellipticity components $e_1$ (e) and $e_2$ (f). Tree rings constitute the dominant source of systematic error, affecting astrometry most significantly.}
    \label{fig:des-results}
\end{figure*}

\vspace{0.15in}
We apply our method to a coadded flat field from a DECam sensor (shown at left in Figure \ref{fig:des-resid}) and obtain the residuals illustrated in the right panel in the same figure. The spatial structure of the fitting residuals is distinct from that of the data, indicating that the fitted model contains nearly all the relevant spatial structure present in the data.

Figure \ref{fig:des-results} presents the impact of sensor effects on 5 key astronomical observables as modeled across a DECam chip using our method. Included are photometry and astrometry, PSF shape, parametrized by PSF size $e_0$, and unnormalized ellipticities $e_1$ and $e_2$ as defined in section \ref{sec:method}. A flat field (smoothed with a nominal PSF kernel to highlight medium-scale structure) is included at left so structure in the systematics maps can be visually compared to structure in the raw data. While there are indeed grid distortions at the single pixel scale in a real sensor, as we showed in section \ref{sec:validation}, the systematic impact of these is negligible for any source that is well-sampled, thus justifying the smoothing of the flat field for visual comparison. 

Tree rings provide the dominant spatial structure in all 5 systematics maps. The impact of the tree rings is most evident in the astrometric offset, where their imprint is observed at the $\sim 10 \text{ mas}$ level. Given this significant impact, DES is currently implementing a catalog-level fix for this using star flats as described in \cite{plazas}. 

The variations in photometry and PSF size correlate with tree rings since the effective area of pixels near tree ring peaks is larger than those near troughs. This constitutes an approximately affine stretching of the local pixel grid near tree ring extrema which is unaccounted for in traditional PSF estimation. The effect on PSF shape parameters agrees with expectation based on Figure \ref{fig:pixshapes}, as $e_1$ is primarily affected in areas where tree ring variations run vertically and horizontally, while $e_2$ is affected only when the tree rings run diagonally. The issue of DECam tree rings impacting PSF behavior is currently being addressed within the DES collaboration.

\subsection{Results from LSST}
\label{sec:lsst}

\begin{figure}[!t]
    \centering
    \includegraphics[width=.7\columnwidth]{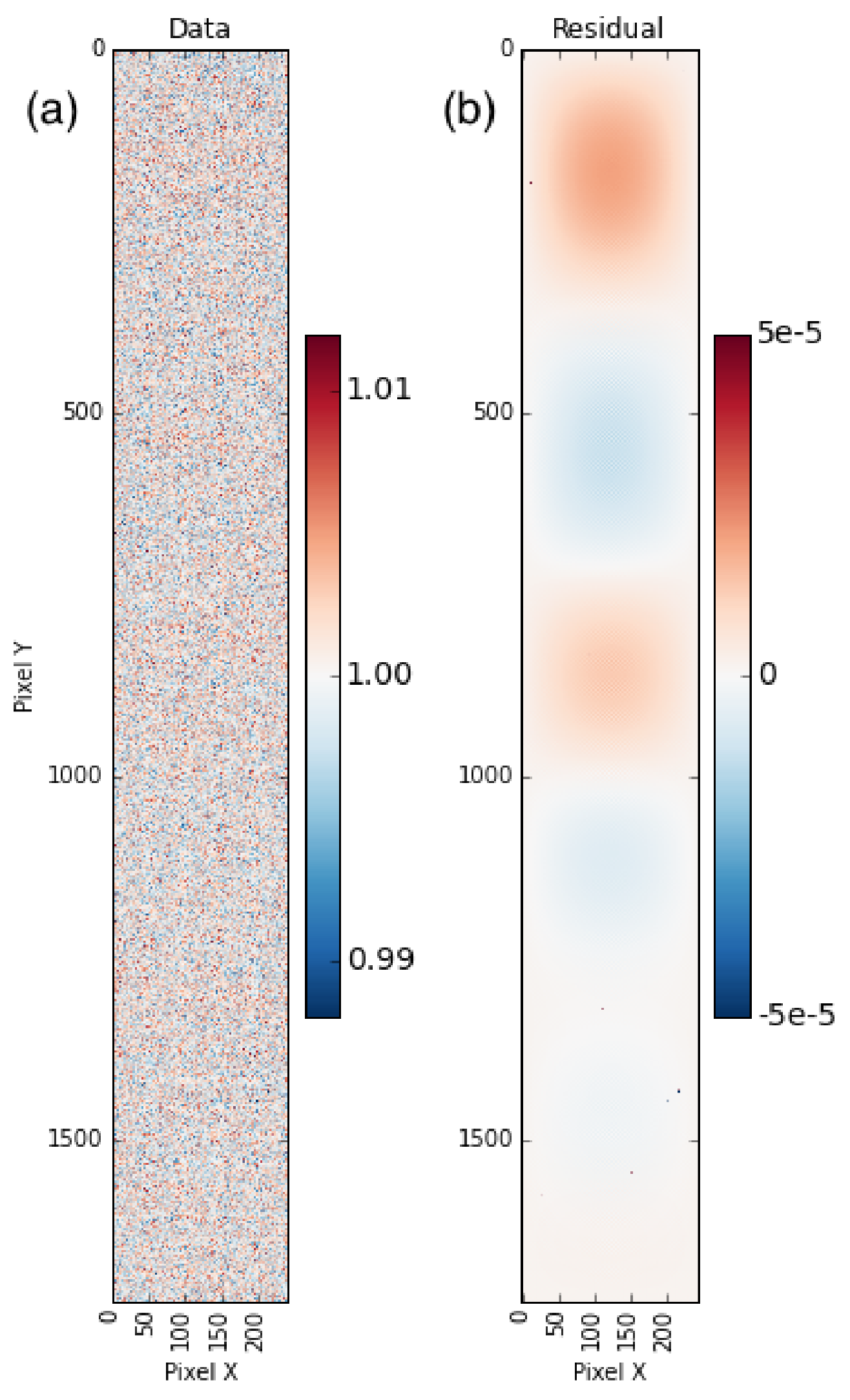}
    \caption{\normalsize A single amplifier segment from a coadd of $\sim 500$ flat field images (a) from an LSST prototype sensor, and the residual map (b) resulting from fitting a pixel grid model as described in section \ref{sec:method}, showing that the model reproduces 99.85\% of the RMS flux variations present in the coadded flat field.}
    \label{fig:residual}
\end{figure}
\vspace{0.1in}

\begin{figure*}[htbp]
    \centering
    \includegraphics[height=3in]{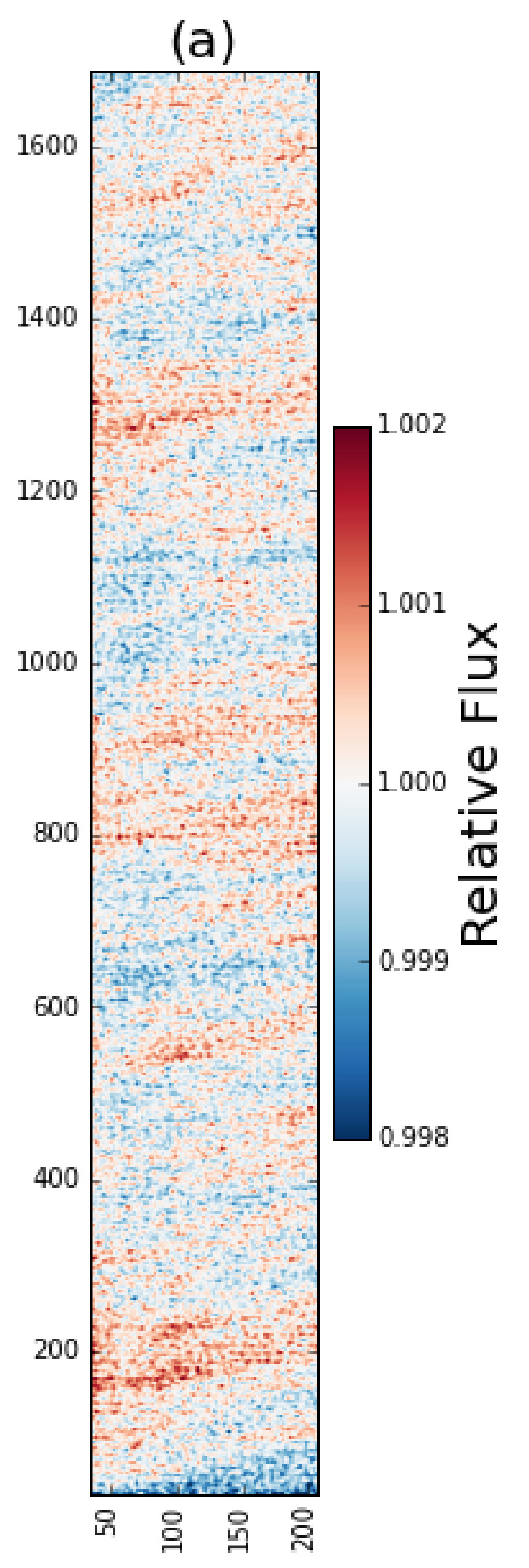}
    \includegraphics[height=3in]{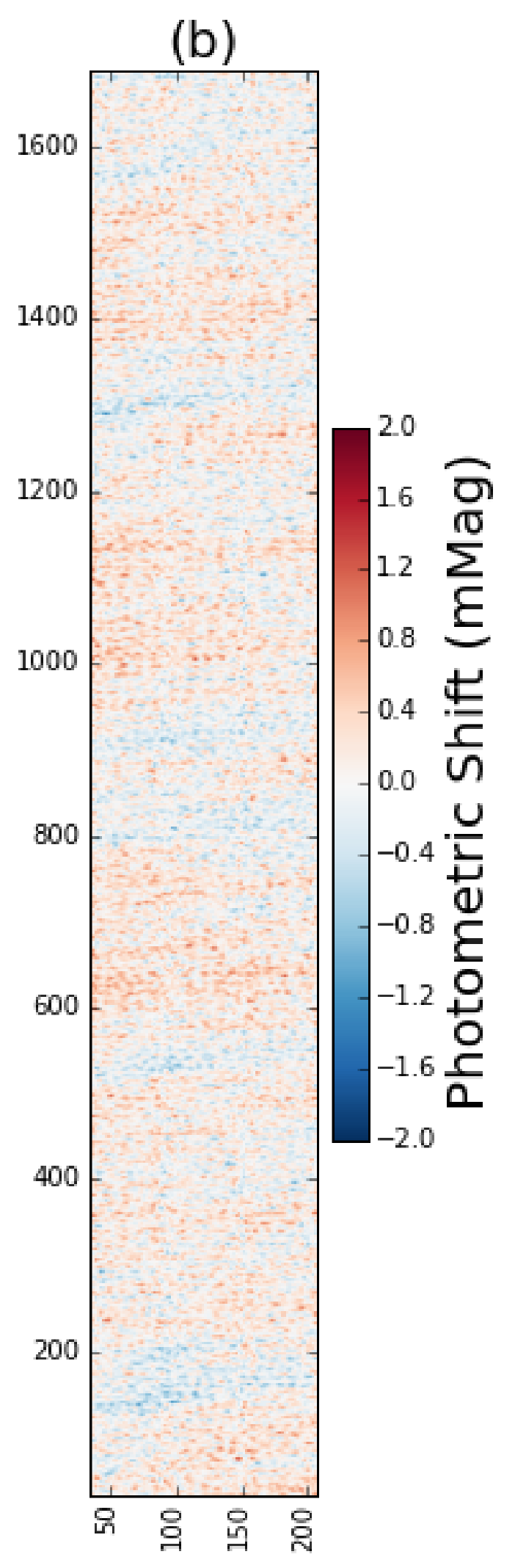}
    \includegraphics[height=3in]{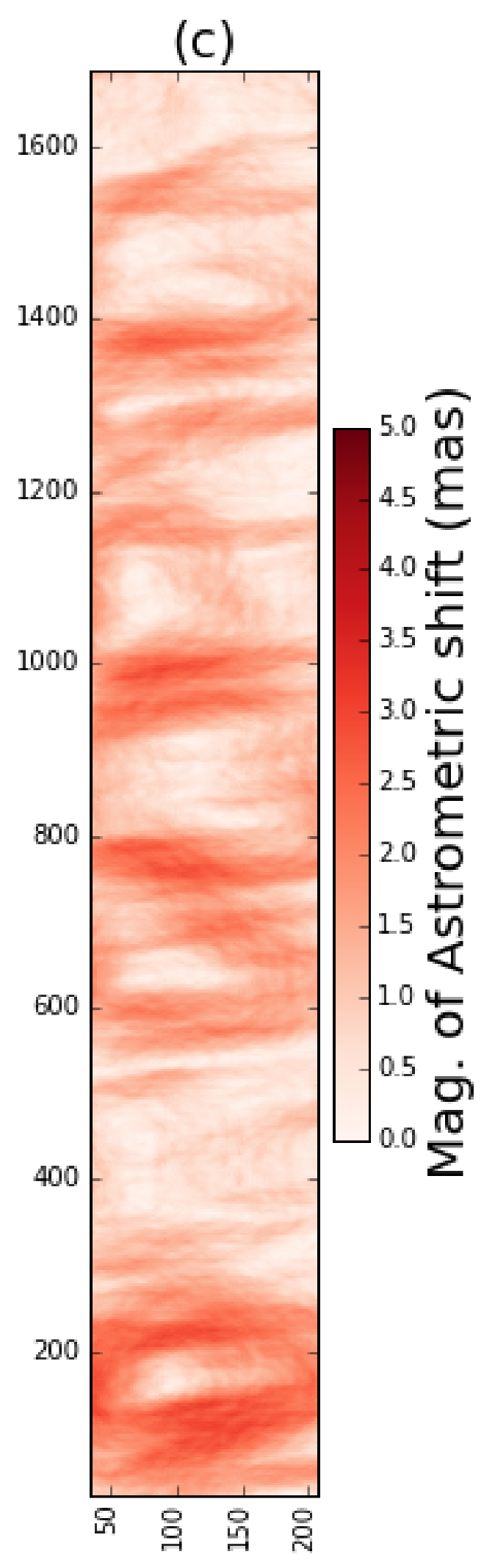}
    \includegraphics[height=3in]{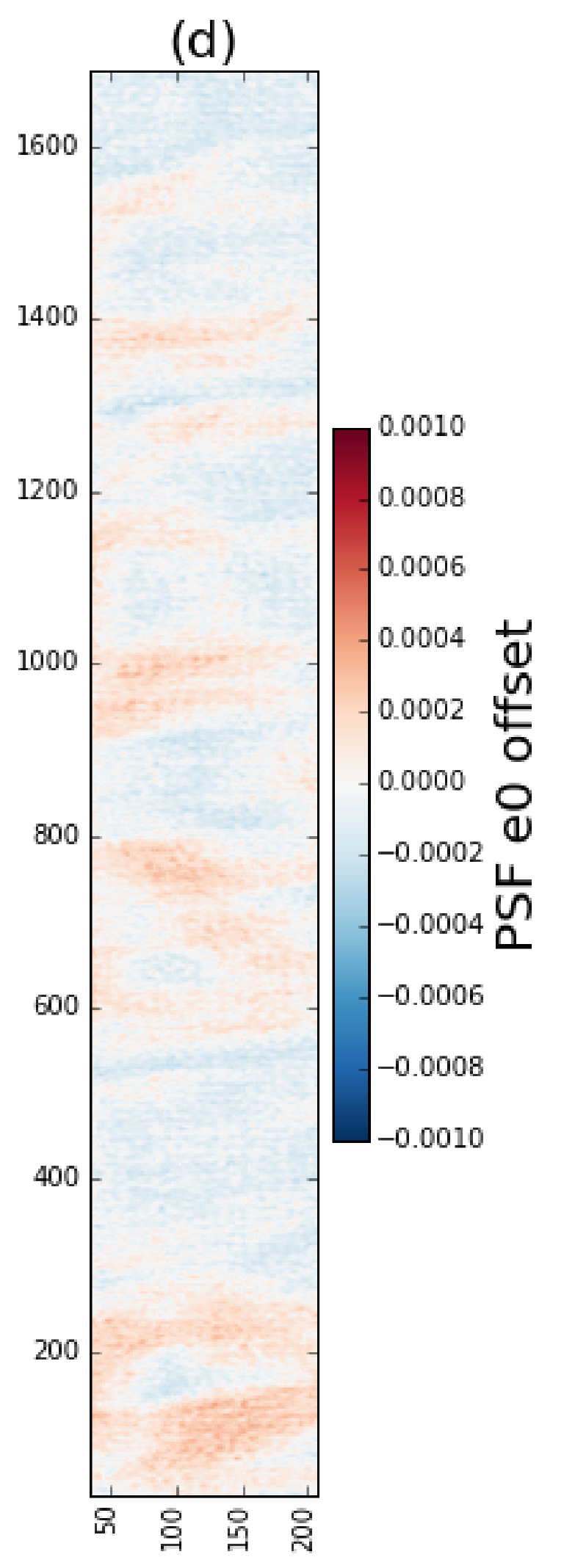}
    \includegraphics[height=3in]{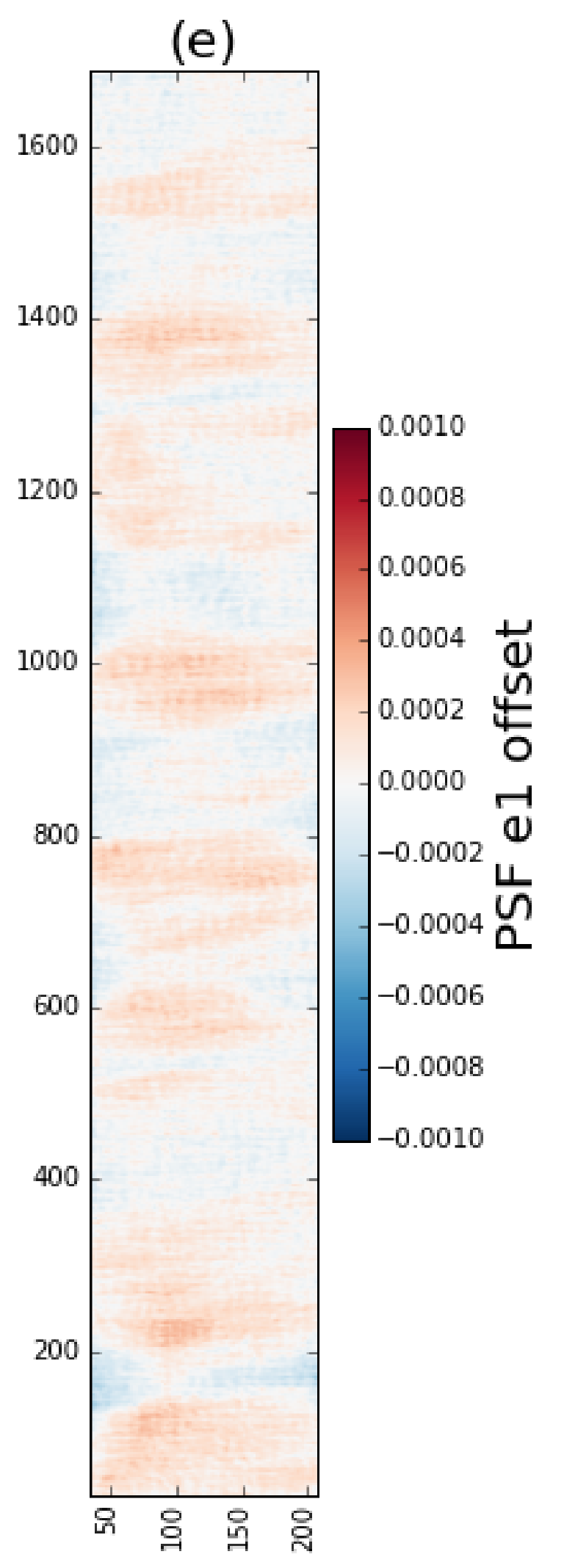}
    \includegraphics[height=3in]{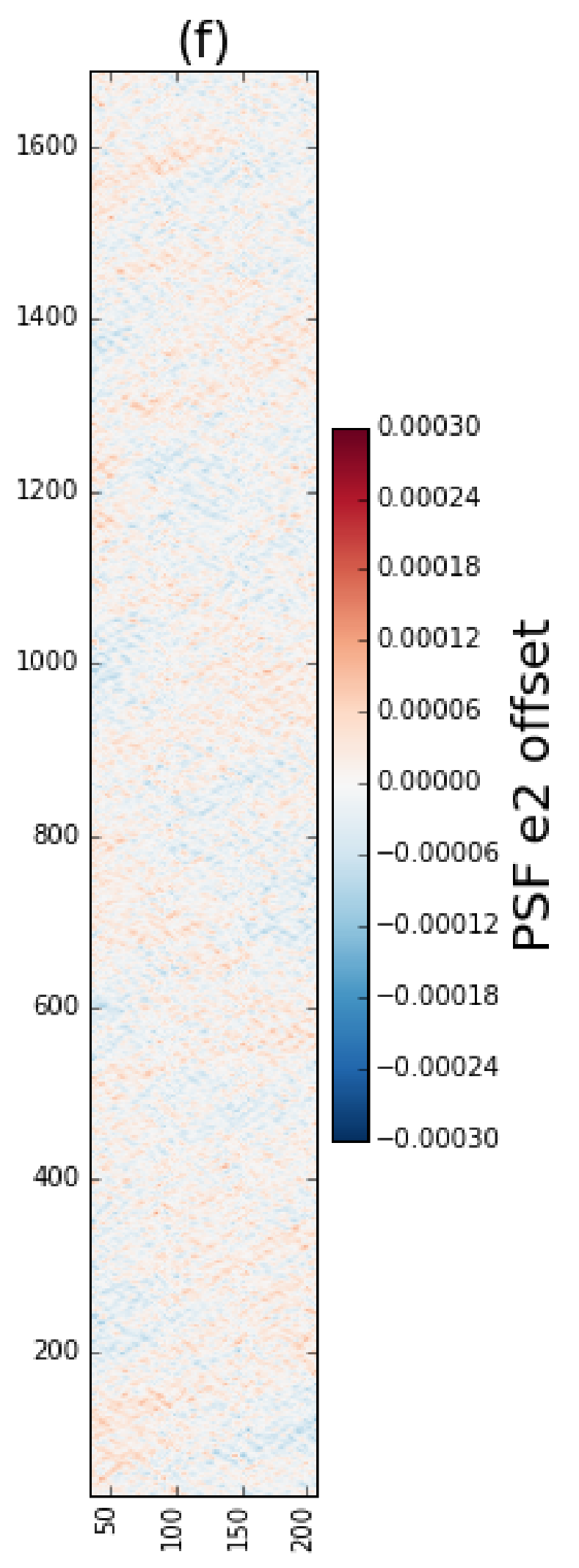}
    \caption{\normalsize Similar to Figure \ref{fig:des-results}, systematics maps from an LSST prototype sensor showing the impact of pixel grid distortions on photometry (b), astrometry (c), PSF size (d), and PSF ellipticity components $e_1$ (e) and $e_2$ (f), with a lightly smoothed flat field (a) for comparison. The observed structure in the systematics maps is genuine, as it is uncorrelated with the residuals shown in Figure \ref{fig:residual}.}
    \label{fig:results}
\end{figure*}

Applying our method to the LSST data described in section \ref{sec:data} yields a fitted grid model that captures 99.85\% of the flux deviations present in the data. Figure \ref{fig:residual} shows the flat field the model is trying to match, along with the map of flux residuals, whose RMS of 0.0006\% is negligible compared with the flat field PRNU of 0.4\%. Again, these residuals are uncorrelated spatially with the target flat field, indicating that the model was successful in picking up spatial structure in the data.

Results from applying the same analysis to an LSST prototype sensor are shown in Figure \ref{fig:results}. None of the systematics residual maps are correlated with the flux residuals from the model fit (cf. Figure \ref{fig:residual}), so the observed structure is not due to a fitting artifact.

Spatial structure is observed in each of the LSST systematics maps, albeit at a much smaller scale than those from DECam. The scatter induced in photometric measurements is well below 1 mMag, while in the astrometric map, a median deviation of about 1 milliarcsecond is observed. The spatial structure visible in the systematics maps is generally observed at locations where the gradient of the coadded flat field is highest. This is where the assumption of a locally-rectilinear (affine) pixel grid is maximally violated.

While photometric and astrometric precision will be of general concern to nearly all users of LSST data, the PSF shape parameters, also affected by the modeled grid distortions, are of primary importance to weak gravitational lensing analyses. As Figure \ref{fig:results} shows, the fitted pixel grid distortions do cause some spatial structure to appear in the maps of $e_0$, $e_1$, and $e_2$. However, since cosmology from weak lensing relies on correlation functions of shape rather than individual measurements, the requirement on PSF shape measurements is expressed as a correlation function, $\xi^{PSF}(e,e)$ as elucidated in \cite{desshear}. For a fixed level of PSF leakage (contamination of galaxy shapes by unremoved PSF), a requirement on PSF shear auto-correlations can be derived; we use the requirements given in \cite{lsstscireq}. Computing the correlation functions of our derived PSF shape errors yields structure-free correlations orders of magnitude below LSST requirements.

\begin{table}[htbp]
    \centering
    \begin{tabular}{lcccc} \hline \hline
    \multirow{2}{*}{Observable}    & Flux & Astrometry & \multirow{2}{*}{$\xi(e_1,e_1)$} & \multirow{2}{*}{$\xi(e_2,e_2)$}  \\
    & (mMag) & (mas) & & \\ \hline
    Shift $\mu$     & 0.08 & 1.2& \multirow{2}{*}{$<10^{-8}$}& \multirow{2}{*}{$<5\times 10^{-10}$}\\
    Scatter $\sigma$ & 0.31 & 0.7 & &\\ \hline
    LSST Req.& 5 & 10 & $2\times10^{-5}$ & $10^{-7}$\\ \hline
    \end{tabular}
    \caption{\normalsize This table shows the photometric and astrometric mean shifts $\mu$ and induced scatter $\sigma$ due to induced by the pixel grid distortions fit by our model. PSF shape errors are quoted as upper bounds on correlation functions. For a nominal sensor with a PRNU of 0.4\%, all three fall well below LSST requirements \citep{lsstscireq}.}
    \label{tab:results}
\end{table}

A summary of results from this analysis is presented in Table \ref{tab:results}. The mean shift $\mu$ refers to the average offset induced by the fitted pixel grid distortions, and the induced scatter $\sigma$ is the scatter in an observable quantity across a sensor chip due to these effects. Overall, this shows that the impact of all static sensor effects taken together, including unknown effects that impact flat fields, are negligible compared to LSST single-exposure requirements (these systematics should be even further suppressed by LSST's many exposures and dithering strategy). This indicates that flat-fielding will remain an appropriate strategy for calibrating LSST images. However, this particular sensor has a PRNU of only 0.4\%, much lower than the LSST CCD requirement of 5\%. We consider the implications of this in the following section.

\section{Discussion}
\label{sec:discussion}

Although we have shown that the impact of static sensor effects on LSST science observables is modest for a sensor with very low PRNU, it is possible that some of the sensors included in the LSST focal plane could have a PRNU approaching 5\%. In this section, we consider the impact of such high PRNU if it is due to pixel grid distortions rather than local QE variation (which would be correctable with typical flat-fielding).

\begin{figure}[htbp]
    \centering
    \includegraphics[width=\columnwidth]{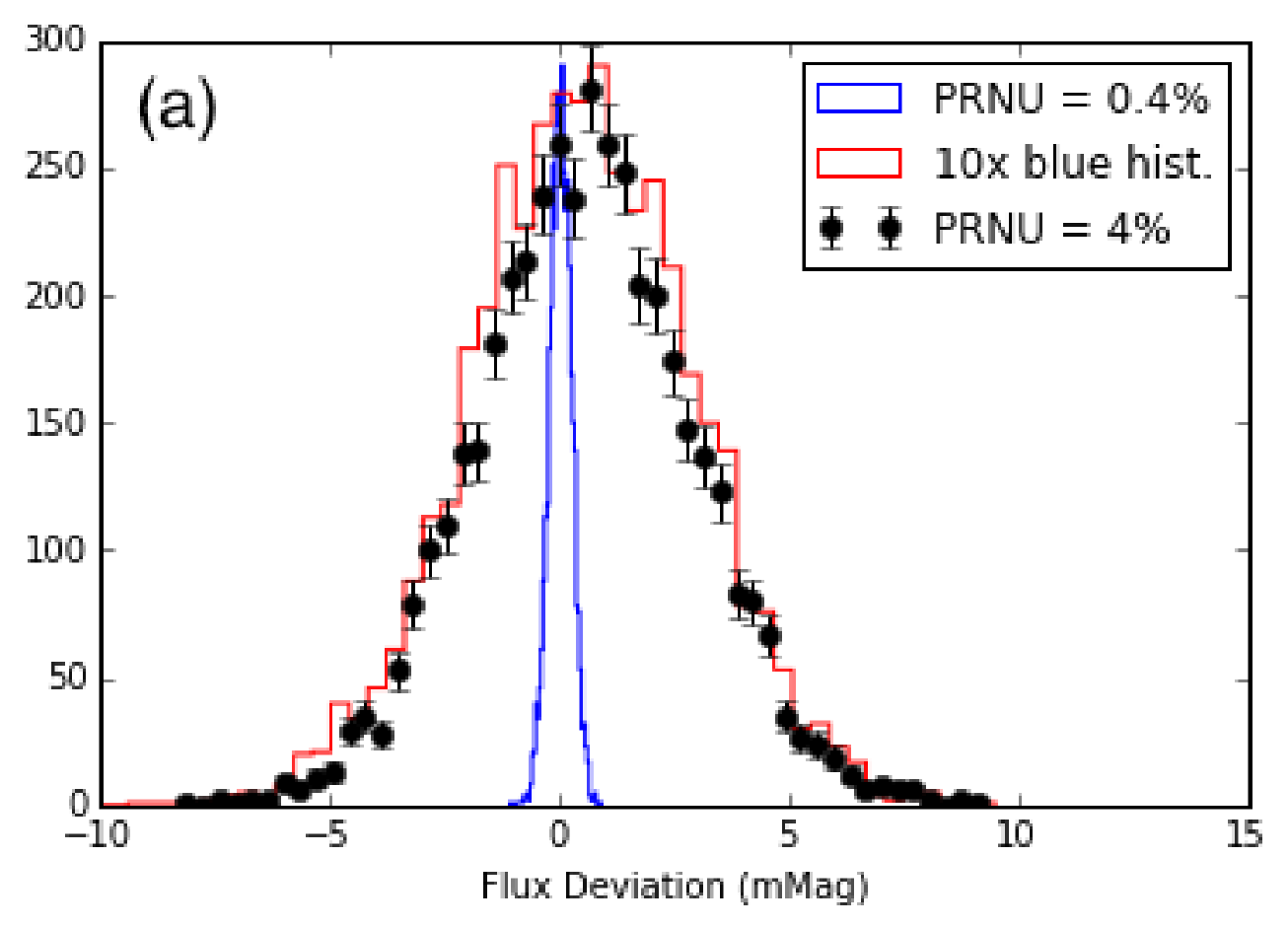}
    \includegraphics[width=\columnwidth]{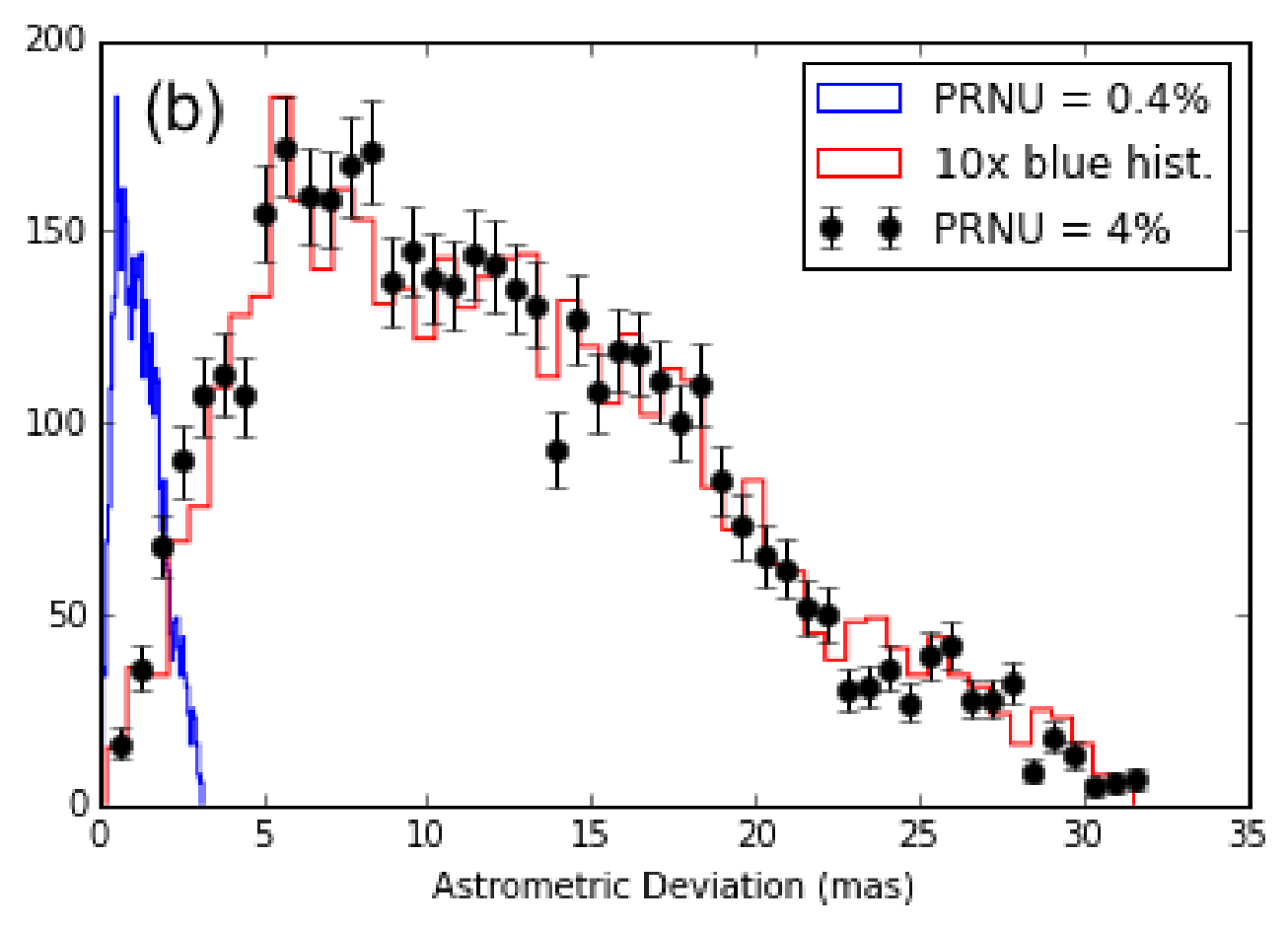}
    \caption{\normalsize The correspondence between photometric (a) and astrometric (b) shifts from the LSST sensor with PRNU=$0.4\%$ scaled up by a factor of 10 and the shifts obtained by depositing PSF profiles on a grid with vertex perturbations scaled by the same factor shows that we remain in the linear distortion regime even for relatively high PRNU.}
    \label{fig:photo_scaleup}
\end{figure}

In Figure \ref{fig:photo_scaleup}, we investigate this by multiplicatively scaling up the vertex displacements from our fitted model by a factor of 10, resulting in a flat field with $\sim 4\%$ PRNU. The photometric and astrometric shifts obtained from this analysis are shown as black points shown in Figure \ref{fig:photo_scaleup}. 

Interestingly, directly scaling up the photometric distortions observed with the original fitted pixel grid (blue), yields the distribution illustrated by the red histogram in Figure \ref{fig:photo_scaleup}. The agreement between these two distributions shows that even for pixel grids near the worst-case LSST PRNU, the systematic errors on science observables respond linearly to increased grid distortions. This linearity allows us to predict the scale of the systematics that might arise from any sensor of moderately-low PRNU. Specifically, the values quoted in Table \ref{tab:results} can be scaled linearly (quadratically for the correlation functions) to other PRNU values as a way to estimate systematic impact of static effects in lower quality sensors. However, given the generality of the framework proposed in this work, it would also make sense to apply it to incoming production sensors to test whether their unique pixel grid structure led to different systematic effects.

\section{Conclusions}
\label{sec:conclusions}

We have presented a method to infer a model of an underlying pixel grid that agnostically models the superposition of all electrostatic (curl-free) static sensor effects present in flat fields, reproducing observed flat field fluxes with very high ($\sim 99\%$) fidelity. We have used these pixel grid models to analyze the impact of sensor systematics on photometry, astrometry, and PSF shape measurements for both DECam and LSST prototype sensors. 

For the DECam sensor, our findings recover behavior of tree rings known from both electrostatic simulation and star flat data, and agree with previous work in finding significant impact of static pixel grid distortions on astrometry and PSF observables, for which independent corrections are currently being developed within the DES collaboration. 

For the LSST prototype, we find that for a sensor with low PRNU ($0.4\%$), the impact all static sensor effects taken together---both known and unknown---on these observables is negligible, but that the impact scales linearly with the PRNU. Therefore, for sensors with PRNU values closer to the specification of $5\%$, additional work to separate and model the effects of local QE variation and pixel grid distortions may be required. 

In cases where a fitted pixel grid model uncovers an unacceptably high level of systematics, or in the case of images with exceptional seeing, where the impact of static pixel grid distortions would be magnified, our method could potentially be used as the basis of a correction to ameliorate them. The primary advantage of such a correction would be that the sampling of systematics maps can be arbitrarily dense, rather than being limited by the finite density of real star flats, for example. A proper correction could require extending this framework to incorporate additional data (for example, flat fields with non-uniform illumination patterns) to disentangle local QE variation from true pixel grid distortions. We leave consideration of such correction algorithms as future work.

Overall, we conclude that for sensors with low PRNU, calibrating LSST images using traditional flat-fielding, despite the presence of varying pixel size, would have negligible impact on LSST science, even in the worst case that all of the PRNU is due to pixel size variation. However, for sensors with PRNU closer to the nominal worst case, more characterization may be required to disentangle the individual contributions of pixel grid distortions and local QE variation to the PRNU to form the basis of a more detailed correction. As long as LSST production sensors have sufficiently low PRNU, we anticipate that this will not be necessary.

\acknowledgments{We'd like to thank Devon Powell for advice in adapting his phase-space integration code, Andres Plazas for providing DECam star flat measurements, and Peter Doherty for providing LSST flat field data. Many thanks to Robert Lupton and Gary Bernstein for their insightful comments on the manuscript. We are grateful for the extraordinary contributions of our CTIO and DECam colleagues in achieving the excellent instrument
and telescope conditions that have made this work possible. This work was performed in part under DOE Contract DE-AC02-76SF00515. This material is based upon work supported by the National Science Foundation Graduate Research Fellowship under Grant No. DGE-114747.}


\bibliographystyle{apj}
\bibliography{references}

\begin{thebibliography}{}
\expandafter\ifx\csname natexlab\endcsname\relax\def\natexlab#1{#1}\fi

\bibitem[{{Antilogus} {et~al.}(2014){Antilogus}, {Astier}, {Doherty},
  {Guyonnet}, \& {Regnault}}]{antilogus}
{Antilogus}, P., {Astier}, P., {Doherty}, P., {Guyonnet}, A., \& {Regnault}, N.
  2014, Journal of Instrumentation, 9, C03048

\bibitem[{{Baumer} \& {Roodman}(2015)}]{me!}
{Baumer}, M.~A., \& {Roodman}, A. 2015, Journal of Instrumentation, 10, C05024

\bibitem[{{Bernstein} {et~al.}(2017){Bernstein}, {Armstrong}, {Plazas},
  {Walker}, {Abbott}, {Allam}, {Bechtol}, {Benoit-L{\'e}vy}, {Brooks}, {Burke},
  {Carnero Rosell}, {Carrasco Kind}, {Carretero}, {Cunha}, {da Costa}, {DePoy},
  {Desai}, {Diehl}, {Eifler}, {Fernandez}, {Fosalba}, {Frieman},
  {Garc{\'{\i}}a-Bellido}, {Gerdes}, {Gruen}, {Gruendl}, {Gschwend},
  {Gutierrez}, {Honscheid}, {James}, {Kent}, {Krause}, {Kuehn}, {Kuropatkin},
  {Li}, {Maia}, {March}, {Marshall}, {Menanteau}, {Miquel}, {Ogando}, {Reil},
  {Roodman}, {Rykoff}, {Sanchez}, {Scarpine}, {Schindler}, {Schubnell},
  {Sevilla-Noarbe}, {Smith}, {Smith}, {Soares-Santos}, {Sobreira}, {Suchyta},
  {Swanson}, \& {Tarle}}]{gary}
{Bernstein}, G.~M., {Armstrong}, R., {Plazas}, A.~A., {et~al.} 2017, ArXiv
  e-prints, arXiv:1703.01679

\bibitem[{{Bradshaw} {et~al.}(2015){Bradshaw}, {Lage}, {Resseguie}, \&
  {Tyson}}]{bradshaw}
{Bradshaw}, A., {Lage}, C., {Resseguie}, E., \& {Tyson}, J.~A. 2015, Journal of
  Instrumentation, 10, C04034

\bibitem[{{Dark Energy Survey Collaboration}(2005)}]{des}
{Dark Energy Survey Collaboration}. 2005, ArXiv Astrophysics e-prints,
  astro-ph/0510346

\bibitem[{{Dark Energy Survey Collaboration}(2016)}]{des2}
---. 2016, \mnras, 460, 1270

\bibitem[{{Flaugher} {et~al.}(2015){Flaugher}, {Diehl}, {Honscheid}, {Abbott},
  {Alvarez}, {Angstadt}, {Annis}, {Antonik}, {Ballester}, {Beaufore},
  {Bernstein}, {Bernstein}, {Bigelow}, {Bonati}, {Boprie}, {Brooks},
  {Buckley-Geer}, {Campa}, {Cardiel-Sas}, {Castander}, {Castilla}, {Cease},
  {Cela-Ruiz}, {Chappa}, {Chi}, {Cooper}, {da Costa}, {Dede}, {Derylo},
  {DePoy}, {de Vicente}, {Doel}, {Drlica-Wagner}, {Eiting}, {Elliott}, {Emes},
  {Estrada}, {Fausti Neto}, {Finley}, {Flores}, {Frieman}, {Gerdes},
  {Gladders}, {Gregory}, {Gutierrez}, {Hao}, {Holland}, {Holm}, {Huffman},
  {Jackson}, {James}, {Jonas}, {Karcher}, {Karliner}, {Kent}, {Kessler},
  {Kozlovsky}, {Kron}, {Kubik}, {Kuehn}, {Kuhlmann}, {Kuk}, {Lahav}, {Lathrop},
  {Lee}, {Levi}, {Lewis}, {Li}, {Mandrichenko}, {Marshall}, {Martinez},
  {Merritt}, {Miquel}, {Mu{\~n}oz}, {Neilsen}, {Nichol}, {Nord}, {Ogando},
  {Olsen}, {Palaio}, {Patton}, {Peoples}, {Plazas}, {Rauch}, {Reil}, {Rheault},
  {Roe}, {Rogers}, {Roodman}, {Sanchez}, {Scarpine}, {Schindler}, {Schmidt},
  {Schmitt}, {Schubnell}, {Schultz}, {Schurter}, {Scott}, {Serrano}, {Shaw},
  {Smith}, {Soares-Santos}, {Stefanik}, {Stuermer}, {Suchyta}, {Sypniewski},
  {Tarle}, {Thaler}, {Tighe}, {Tran}, {Tucker}, {Walker}, {Wang}, {Watson},
  {Weaverdyck}, {Wester}, {Woods}, {Yanny}, \& {DES Collaboration}}]{decam}
{Flaugher}, B., {Diehl}, H.~T., {Honscheid}, K., {et~al.} 2015, \aj, 150, 150

\bibitem[{{Gruen} {et~al.}(2015){Gruen}, {Bernstein}, {Jarvis}, {Rowe},
  {Vikram}, {Plazas}, \& {Seitz}}]{gruen}
{Gruen}, D., {Bernstein}, G.~M., {Jarvis}, M., {et~al.} 2015, Journal of
  Instrumentation, 10, C05032

\bibitem[{{Guyonnet} {et~al.}(2015){Guyonnet}, {Astier}, {Antilogus},
  {Regnault}, \& {Doherty}}]{guyonnet}
{Guyonnet}, A., {Astier}, P., {Antilogus}, P., {Regnault}, N., \& {Doherty}, P.
  2015, \aap, 575, A41

\bibitem[{{Hirata} \& {Seljak}(2003)}]{hsm}
{Hirata}, C., \& {Seljak}, U. 2003, \mnras, 343, 459

\bibitem[{{Holland} {et~al.}(2014){Holland}, {Bebek}, {Kolbe}, \&
  {Lee}}]{holland}
{Holland}, S.~E., {Bebek}, C.~J., {Kolbe}, W.~F., \& {Lee}, J.~S. 2014, Journal
  of Instrumentation, 9, C03057

\bibitem[{Hunter(2007)}]{matplotlib}
Hunter, J.~D. 2007, Computing In Science \& Engineering, 9, 90

\bibitem[{{Jarvis} {et~al.}(2016){Jarvis}, {Sheldon}, {Zuntz}, {Kacprzak},
  {Bridle}, {Amara}, {Armstrong}, {Becker}, {Bernstein}, {Bonnett}, {Chang},
  {Das}, {Dietrich}, {Drlica-Wagner}, {Eifler}, {Gangkofner}, {Gruen},
  {Hirsch}, {Huff}, {Jain}, {Kent}, {Kirk}, {MacCrann}, {Melchior}, {Plazas},
  {Refregier}, {Rowe}, {Rykoff}, {Samuroff}, {S{\'a}nchez}, {Suchyta},
  {Troxel}, {Vikram}, {Abbott}, {Abdalla}, {Allam}, {Annis}, {Benoit-L{\'e}vy},
  {Bertin}, {Brooks}, {Buckley-Geer}, {Burke}, {Capozzi}, {Rosell}, {Kind},
  {Carretero}, {Castander}, {Clampitt}, {Crocce}, {Cunha}, {D'Andrea}, {da
  Costa}, {DePoy}, {Desai}, {Diehl}, {Doel}, {Neto}, {Flaugher}, {Fosalba},
  {Frieman}, {Gaztanaga}, {Gerdes}, {Gruendl}, {Gutierrez}, {Honscheid},
  {James}, {Kuehn}, {Kuropatkin}, {Lahav}, {Li}, {Lima}, {March}, {Martini},
  {Miquel}, {Mohr}, {Neilsen}, {Nord}, {Ogando}, {Reil}, {Romer}, {Roodman},
  {Sako}, {Sanchez}, {Scarpine}, {Schubnell}, {Sevilla-Noarbe}, {Smith},
  {Soares-Santos}, {Sobreira}, {Swanson}, {Tarle}, {Thaler}, {Thomas},
  {Walker}, \& {Wechsler}}]{desshear}
{Jarvis}, M., {Sheldon}, E., {Zuntz}, J., {et~al.} 2016, \mnras, 460, 2245

\bibitem[{{LSST Science Collaboration}(2009)}]{scibook}
{LSST Science Collaboration}. 2009, ArXiv e-prints, arXiv:0912.0201

\bibitem[{{LSST Science Council}(2011)}]{lsstscireq}
{LSST Science Council}. 2011, {LSST System Science Requirements Document},
  LPM-17

\bibitem[{{Lupton}(2014)}]{lupton}
{Lupton}, R.~H. 2014, Journal of Instrumentation, 9, C04023

\bibitem[{McKinney(2010)}]{pandas}
McKinney, W. 2010, in Proceedings of the 9th Python in Science Conference, ed.
  S.~van~der Walt \& J.~Millman, 51 -- 56

\bibitem[{{O'Connor}(2014)}]{oconnor}
{O'Connor}, P. 2014, Journal of Instrumentation, 9, C03033

\bibitem[{{Okura} {et~al.}(2016){Okura}, {Petri}, {May}, {Plazas}, \&
  {Tamagawa}}]{newyuki}
{Okura}, Y., {Petri}, A., {May}, M., {Plazas}, A.~A., \& {Tamagawa}, T. 2016,
  \apj, 825, 61

\bibitem[{{Okura} {et~al.}(2015){Okura}, {Plazas}, {May}, \&
  {Tamagawa}}]{okuraTR}
{Okura}, Y., {Plazas}, A.~A., {May}, M., \& {Tamagawa}, T. 2015, Journal of
  Instrumentation, 10, C08010

\bibitem[{Perez \& Granger(2007)}]{ipython}
Perez, F., \& Granger, B.~E. 2007, Computing in Science \& Engineering, 9, 21

\bibitem[{Plazas {et~al.}(2014)Plazas, Bernstein, \& Sheldon}]{plazas}
Plazas, A.~A., Bernstein, G.~M., \& Sheldon, E.~S. 2014, Publications of the
  Astronomical Society of the Pacific, 126, 750

\bibitem[{{Plazas} {et~al.}(2017){Plazas}, {Shapiro}, {Smith}, {Rhodes}, \&
  {Huff}}]{plazas2}
{Plazas}, A.~A., {Shapiro}, C., {Smith}, R., {Rhodes}, J., \& {Huff}, E. 2017,
  Journal of Instrumentation, 12, C04009

\bibitem[{{Powell} \& {Abel}(2015)}]{devon}
{Powell}, D., \& {Abel}, T. 2015, Journal of Computational Physics, 297, 340

\bibitem[{{Rasmussen}(2015)}]{andy}
{Rasmussen}, A. 2015, Journal of Instrumentation, 10, C05028

\bibitem[{{Rowe} {et~al.}(2015){Rowe}, {Jarvis}, {Mandelbaum}, {Bernstein},
  {Bosch}, {Simet}, {Meyers}, {Kacprzak}, {Nakajima}, {Zuntz}, {Miyatake},
  {Dietrich}, {Armstrong}, {Melchior}, \& {Gill}}]{galsim}
{Rowe}, B.~T.~P., {Jarvis}, M., {Mandelbaum}, R., {et~al.} 2015, Astronomy and
  Computing, 10, 121

\bibitem[{{Smith} \& {Rahmer}(2008)}]{smith}
{Smith}, R.~M., \& {Rahmer}, G. 2008, in \procspie, Vol. 7021, High Energy,
  Optical, and Infrared Detectors for Astronomy III, 70212A

\bibitem[{{Stubbs}(2014)}]{stubbs}
{Stubbs}, C.~W. 2014, Journal of Instrumentation, 9, C03032

\bibitem[{van~der Walt {et~al.}(2011)van~der Walt, Colbert, \&
  Varoquaux}]{numpy}
van~der Walt, S., Colbert, S.~C., \& Varoquaux, G. 2011, Computing in Science
  \& Engineering, 13, 22

\end{thebibliography}

\end{document}